\documentclass[journal]{IEEEtran}

\usepackage{graphics} 
\usepackage{mathptmx} 
\usepackage{times} 
\usepackage{amsmath} 
\usepackage{amsfonts}  
\usepackage{enumerate} 
\usepackage{subfig}
\usepackage{makecell} 
\usepackage{multirow}
\usepackage{mathtools}
\usepackage[english, onelanguage,ruled,vlined, longend, titlenotnumbered]{algorithm2e}

\newtheorem{rem}{Remark}
\newtheorem{lem}{Lemma}

\newcommand{\RR}{{\mathbb R}}

\DeclareMathOperator*{\argmin}{argmin}
\DeclareMathOperator*{\argmax}{argmax}

\usepackage{pgfplots,pgfplotstable}
\usepackage{tikz}
\usetikzlibrary[patterns]
\pgfplotsset{compat=1.12}
\usetikzlibrary{plotmarks}
\usetikzlibrary{arrows,shapes,positioning}
\usetikzlibrary{calc,trees,positioning,arrows,chains,shapes.geometric,%
	decorations.pathreplacing,decorations.pathmorphing,shapes,%
	matrix,shapes.symbols}
\usetikzlibrary{decorations.markings}

\usepackage{setspace}

\usepackage{hyperref}
\hypersetup{hidelinks}

\begin{document}
	\title{Factor graph based smoothing without matrix inversion for highly precise localization}
	\author{Paul \textsc{Chauchat},
			Axel  \textsc{Barrau} and
			Silv\`ere \textsc{Bonnabel}%
		\thanks{Paul Chauchat is with ISAE-Supa\'ero, University of Toulouse, 31055 Toulouse, France (e-mail: paul.chauchat@isae-supaero.fr)}%
		\thanks{Silv\`ere Bonnabel is with MINES ParisTech, PSL Reasearch University, Centre for Robotics, 60 bd Saint-Michel, 75006 Paris, France (e-mail: silvere.bonnabel@mines-paristech.fr)}%
		\thanks{Axel Barrau is with SAFRAN TECH, Groupe Safran, Rue des Jeunes Bois - Ch\^ateaufort, 78772 Magny Les Hameaux CEDEX, France (e-mail: axel.barrau@safrangroup.com)}%
	}
	
	\maketitle
	\begin{abstract}
		We consider the problem of localizing a manned, semi-autonomous, or autonomous vehicle in the environment using information coming from the vehicle's sensors, a problem known as navigation or simultaneous localization and mapping (SLAM) depending on the context. To infer knowledge from sensors' measurements, while drawing on a priori knowledge about the vehicle's dynamics, modern approaches solve an optimization problem to compute the most likely trajectory given all past observations, an approach known as smoothing.   
		Improving smoothing solvers is an active field of research in the SLAM community. Most work is focused on reducing computation load by inverting the involved linear system while preserving its sparsity. The present paper raises an issue which, to the knowledge of the authors, has not been addressed yet: standard smoothing solvers require explicitly using the inverse of sensor noise covariance matrices. This means the parameters that reflect the noise magnitude must be sufficiently large for the smoother to properly function. When  matrices are close to singular, which is the case when using high precision modern inertial measurement units (IMU), numerical issues necessarily arise, especially with 32-bits implementation demanded by most industrial aerospace applications.  We discuss these issues and propose a  solution that builds upon the Kalman filter to improve  smoothing algorithms. We then leverage the results to devise a localization algorithm based on fusion of IMU and vision sensors. Successful real experiments using an actual car equipped with  a tactical grade high performance IMU and a LiDAR illustrate the relevance of the approach to the field of autonomous vehicles.
	\end{abstract}
	\section{INTRODUCTION}

	State estimation is fundamental, notably for control purposes. This is particularly true for autonomous systems, such as autonomous cars. Prevailing approaches to the state estimation problem in robotics, and for inertial navigation in aerospace engineering, explicitly model sensor uncertainties, due to noise and bias, using Gaussian random variables, and then seek to compute the   maximum a posteriori (MAP) state, that is the most likely state in the light of all measurements while drawing on the vehicle's dynamics. Such approaches allow combining sensor measurements optimally based on their confidence levels, as quantified by their covariance, and also allow the estimator to convey a degree of uncertainty associated to its own estimate, which may prove critical for high-level planning and low-level control of autonomous vehicles.

	Simultaneous Localisation and Mapping (SLAM)   has received tremendous attention in the robotics community over the past two decades. With a probabilistic approach, the corresponding mathematical problem may be formulated  as a nonlinear estimation problem, that has served as a benchmark for recent developments in state estimation.   The historical estimator is the extended Kalman filter, which is still state of the art of inertial navigation, but exhibits huge caveats for SLAM, because of its numerical complexity, and especially its inherent inconsistency, that prompted the use of  particle filters instead \cite{dissanayake2001solution, julier2001counter, montemerlo2003fastslam, durrant2006simultaneous}. More recently, non-linear observers and filters were reintroduced for SLAM through the use of geometric frameworks, and Lie groups in particular \cite{brossard2019symmetries, mahony2017geometric, wang2018globally, wang2018geometric}. However, the progresses of vision sensors and vision algorithms over the past years, and their ubiquity in mobile robotics and autonomous driving, have led the community to replace particle filters with optimization based algorithms  being able to return the solution to the MAP problem. Such algorithms have achieved major successes for fusion of vision and inertial measurements, see for instance,  \cite{cadena2016past, stachniss2016simultaneous}. They have also recently been used in GNSS applications \cite{zhao2014differential}, adapting them in various ways to accomodate outliers in this context \cite{pfeifer2018robust, aghapour2018performance, rahman2018outlier}.  They fall into the framework of ``smoothing'' where one seeks to compute the most likely entire past trajectory given sensors measurements. From a theoretical point of view, the nonlinear optimization smoothing problem has been known for decades, but solving it has long proved intractable. 
	
	Smoothing is formulated as a non-linear optimization problem which is  solved through iterative methods, the most popular being the Gauss-Newton (GN), Levenberg-Marquardt, Conjugate Gradient Descent or other trust-region algorithms \cite{dellaert2006square, kaess2012iSAM2, liu2018ice-ba, rosen2014RISE, kuemmerle2011g2o, ceres-solver} methods. Extension to non-gaussian models, also based on successive least-squares problems, were developped \cite{rosen2013robust, pfeifer2018robust, pfeifer2019E-M}. To come up with tractable algorithms, these works heavily exploit the tight link that exists between the MAP problem and factor graphs, and in particular sparse graph structures \cite{dellaert2017factor}.
	
	In the present paper, a  shortcoming of those methods  that has gone unnoticed yet to our best knowledge is raised. Indeed, all the mentioned   formulations are based explicitly on the information matrix of each factor -  the inverse of the covariance matrix -  whose eigenvalues tend to increase as the sensors improve. But a large discrepancy between the most accurate and least accurate sensor leads to ill conditioned information matrices.  This may also impact linearization errors \cite{forster2016preintegration, chauchat2018invariant}. Numerical issues may thus  degrade the solvers' computed solution, and in turn the state estimate accuracy  and consistency. 
	
	As the cost of high performance inertial measurement units (IMU) keeps decreasing, and techological progresses are constantly made,  one may anticipate that tactical grade IMUs will be used in an increasing number of autonomous systems. Their precision is so high that the resulting process noise is close to negligible as compared to other sensors. This prompts the need for solvers being robust to ill conditioning of the information matrix. 
	Using the square-root approach of the information matrix reduces the impact of conditioning \cite{dellaert2006square, maybeck1982stochastic}. Numerical issues related to it depend on the precision of the hardware used, though. Indeed, single precision implementations will much more suffer. This is noticeable as  there is a need to speed up computations for real-time applications, for which single precision is generally considered \cite{wu2015square, chiu2013robust}, and most industrial aerospace computers still use single precision (32 bits or less) \cite{farrell2008aided}. Besides, navigation methods relying on machine learning to improve some of their bricks \cite{brossard2019learning, brossard2019RINS-W}, or mimicking Kalman filters \cite{haarnoja2016backprop}, are on the rise, and some recent methods even rely on Gauss-Newton as one of their bricks \cite{ma2019scene_flow}. They also could benefit from single precision algorithms to speed up their training phase.
	
	In the present paper,  we propose methods to carry out a broad class of non-linear optimization problems while avoiding issues related to ill conditioning, by deriving new exact ways of solving of the linear least-squares problems they induce. More precisely, we   derive two novel linear solvers. First a first simple, robust, batch algorithm for solving least-squares based on a relevant decomposition of the problem. Then, we leverage the correspondence between Bayesian inference and least-squares in the linear case to derive a new sparsity-preserving least-squares solver. To this end, we combine Stochastic Cloning (SC) \cite{mourikis2007sc-kf} and the Backward Information Forward Marginal (BIFM) \cite{loeliger2016sparsity, wadehn2016square} version of the Kalman smoother.  Our starting point was the finding that the Kalman filter never requires to invert process noise covariance matrices, and leads to relatively cheap computation cost,   which allowed simplified implementations on hardware with precision as low as 8-bits processors \cite{martin2010low_cost}. 
	
	These solvers are leveraged to devise a novel smoothing algorithm for localization based on IMU, vision, and additional sensors such as GNSS.  The algorithm is successfully implemented on an actual car equipped with a high performance IMU and a LiDAR, and real experiments illustrate the interest of the approach in the context of highly precise motion sensors. 
	
	The paper is organized  as follows. In Section \ref{sec:smoothing} we summarize the modern factor graph based approaches to the smoothing problem of SLAM, and the issues that arise  when propagation covariances approach zero. We also propose our first (``robust batch'') solver. In Section \ref{BIFM:secc}, we recall how the Kalman smoother may serve as a solver for linear least squares. However, in its standard form it does not accomodate vision measurements. To remedy this issue, we propose our second robust and sparsity-preserving solver, called ``SC-BIFM'', in Section \ref{sec:new_solver}. A simple numerical illustration of the interest of the method is presented in Section \ref{sec:toy_example}. Finally, we demonstrate the validity of our approach in real world experiments in Section \ref{sec:data_joint_lab}, in which the standard methods actually fail in single precision, while our solvers prove robust, leading to a novel accurate visual inertial localization algorithm in the presence of highly precise inertial sensors.
	
	\subsection{Relation to previous literature}
	Most of the smoothing solvers use the information form of the MAP problem, in order to take advantage of its sparsity \cite{dellaert2006square, liu2018ice-ba, ceres-solver}. It has been  long recognised that using the square-root, for instance the Cholesky factorisation of the information matrix, leads to algorithms which are numerically more stable, and even enabled implementations with single precision on mobile devices \cite{wu2015square}. This led to highly efficient incremental implementations \cite{kaess2012iSAM2, liu2018ice-ba}. However, even square-root formulations are inherently limited by the inverses of the covariances appearing in the formulas, and there is no \emph{transition}  between zero covariances, which represent hard constraints, and numerically invertible covariances. This issue has been addressed in the signal processing community by devising new formulations of the Kalman smoother. Additionally to the well-known Modified-Bryson-Frazier Smoother \cite{bierman1977factorization}, a new alternative has been derived, the so-called Backward Information Forward Marginals (BIFM) Smoother, which also avoids covariance inverting \cite{loeliger2016sparsity, wadehn2016square}. Nevertheless, they were only expressed for acyclic graphs (i.e., without relative measurements other than the propagation, such as loop closures), which only cover a small part of the existing factor graphs. We introduce here a more general smoother, able to solve a broader class of linear least-squares, and thus applies to the non-linear optimization problems at hand in navigation, by applying the stochastic cloning method of \cite{mourikis2007sc-kf} to the BIFM.
	\section{Factor graphs and MAP}\label{sec:smoothing}

	Assume we want to track the state of a system  (say, a vehicle) equipped with an IMU and/or wheel odometry, as well as a set of sensors such as cameras, LiDARs, GNSS or acoustic positioning system for UAVs. The current formulation of SLAM uses the formalism of factor graphs to cast the maximum likelihood   estimation problem as a nonlinear optimization problem. Throughout the article, we will not consider landmark-based SLAM, but rather ``pose-graph SLAM'', where vision sensors identify loop closures and provide relative poses between key frames, triggering the back-end optimization run.  Moreover, and contrary to ``pure'' SLAM, we allow absolute sensors such as GNSS to provide information about the vehicle, and refer to SLAM as an extended multi-sensor fusion problem, mainly for localization and navigation purposes. 
	
	\subsection{MAP formulation of the SLAM problem}
	
	\subsubsection{The state} 
	
	The state  is a set of all variables that characterize the vehicle and its sensors, typically its orientation, position, velocity, and all parameters of interest such as IMU biases, camera to IMU transformation parameters.  The state at time $k$ is denoted $X_k$. In landmark-based SLAM, the state also includes the position of the landmarks in the environment.   Without loss of generality the state is assumed to evolve through a discrete-time dynamical model  
	\begin{equation}
	\label{eq::prop}
	X_{k+1} = f(X_k,u_k,w_k)
	\end{equation}
	with $w_k$ a random variable (noise) encoding model uncertainties, and $u_k$ a measurement typically obtained by wheel speed sensors or the IMU.  Generally, the initial uncertainty about the state $X_0$ is assumed Gaussian, that is, $X_0\sim\mathcal N(\hat X_0,P_0)$. In the smoothing approach, one is interested in the entire trajectory,   denoted \begin{equation}
	\label{eq::traj}\chi:=(X_0,X_1,\cdots,X_N).\end{equation} 
	
	\subsubsection{The observations}
	\label{sec:observations}
	
	We are given a set of measurements $Z_1,Z_2,\cdots,Z_K$ of the form $Z_k=h_k(X_{k_1}, \ldots, X_{k_p},V_k)$ where each state is in $X_1,X_2,\cdots,X_k$,   $h_k$ is a known measurement function, and $V_k$ is a random variable   encoding measurements uncertainties that stem from sensors' imperfections (i.e., observations noise). The measurement function of each sensor is fixed, the subscript $k$ indicates that different sensors may be present. In the present paper centered on  inertial navigation, visual inertial odometry, and inertial pose-graph SLAM, we will consider   two types of measurements: \begin{itemize}
		\item Observations that are made at time $k$ and involve only the state at time $k$ (unary observations):
		\begin{equation}
		\label{eq::obs}
		Z_k = h_k(X_k,V_k)
		\end{equation}
		\item Observations at time $k$ that involve a pair of states, that is,
		\begin{equation}
		\label{eq::obs2}
		Z_{ik} = h_{k}(X_i,X_k,V_{k}),\quad i<k
		\end{equation}
		as is the case in pose graph SLAM. These measurements may be related to the computation of relative transformations between poses at different times, typically using stereo cameras or LiDARs. 
		They can be used as either odometry to propagate the state, or as ``loop closures'' \cite{durrant2006simultaneous}.
	\end{itemize}
	For simplicity of notation we   let $Z_k=h_k(\chi,V_k)$ denote both types of observations \eqref{eq::obs} and \eqref{eq::obs2} at time $k$. This is not a problem as $X_k,X_i\in\chi$. Finally $Z=(Z_1,\cdots,Z_N)$ denotes all available measurements. 
	\subsection{Maximum a posteriori (MAP) estimation}
	As only $Z$ is observed, the best one can achieve is to  compute the most likely state  trajectory $\chi$ given all the measurements obtained, that is,  maximum a posteriori (MAP) estimation. Mathematically, one seeks:
	\begin{equation}
	\label{eq::obs222}
	\chi^*=\argmax_\chi p(\chi\mid Z)=\argmax_\chi p(Z\mid \chi )p(\chi ).
	\end{equation}
	Assume the dynamical model \eqref{eq::prop} can be inverted with respect to the noise variable, that is, there exists a function $\phi$ such that
	$$ 
	X_{k+1} = f(X_k,u_k,w_k) \iff w_k = \phi(X_{k+1},X_k,u_k),$$  and similarly $$Z_k = h_k(\chi,V_k) \iff V_k = \psi_k(\chi,Z).$$The latter quantities define \emph{factors} that may be viewed as constraints between variables.  Indeed assuming the noises are independent variables with distribution $w_k\sim\mathcal N(0,Q_k)$, $V_k\sim\mathcal N(0,R_k)$, standard computations show the log-likelihood of the posterior writes 
	\begin{equation}
	\begin{aligned}
	\label{eq::log-likelihood}
	L(\chi) :&= -\log p(\chi\mid Z)=||X_0 - \hat{X}_0||^2_{P_0} \\&+ \sum_{k=0}^{N-1} || \phi(X_{k+1},X_k,u_k) ||^2_{Q_k}   + \sum_{k=1}^N || \psi_k (\chi,Z_k) ||^2_{R_k}. 
	\end{aligned}\end{equation}\emph{up to an additive constant} independent from $\chi$, where $\|e\|_P^2 = e^T P^{-1} e$ is the Mahalanobis norm of the vector $e$ given the covariance $P$. From \eqref{eq::obs222} we see that:
	\begin{equation}
	\label{eq::argmin}
	\chi^* = (X_0^*, \cdots, X_N^*) = \argmin_{\chi} L(\chi).
	\end{equation}
	
	\subsection{Factor graph interpretation}
	\begin{figure}
		\centering
		\begin{tikzpicture}[node distance = 1.5cm]
		\tikzset{Variable/.style = {shape          = circle,
				fill           = blue!30,
				text           = black,
				inner sep      = 2pt,
				outer sep      = 0pt,
				minimum size   = 20 pt}}
		\tikzset{BinaryFactor/.style   = {thick,
		}}
		\tikzset{LabelIMU/.style =   {draw,
				fill           = black}}
		\tikzset{LabelLidar/.style =   {draw,
				fill           = yellow}}
		\tikzset{GPS/.style = {shape          = circle,
				fill           = red,
				text           = black,
				inner sep      = 2pt,
				outer sep      = 0pt,
				minimum size   = 5 pt}}
		\tikzset{Prior/.style = {shape          = diamond,
				fill           = blue,
				text           = black,
				inner sep      = 2pt,
				outer sep      = 0pt,
				minimum size   = 5 pt}}
		
		\node (X0) [Variable] {$X_0$};
		\node (X1) [Variable,right of=X0] {$X_1$};
		\node (X2) [Variable,right of=X1] {$X_2$};
		\node (X3) [Variable,right of=X2] {$X_3$};
		\node (X4) [Variable,right of=X3] {$X_4$};
		
		\node (GPS) [GPS, above = 0.5cm of X4] {};
		\node (Prior) [Prior, left = 0.5cm of X0] {};
		
		\draw[BinaryFactor] (X0) to node[LabelIMU]{} (X1);
		\draw[BinaryFactor] (X1) to node[LabelIMU]{} (X2);
		\draw[BinaryFactor] (X2) to node[LabelIMU]{} (X3);
		\draw[BinaryFactor] (X3) to node[LabelIMU]{} (X4);
		
		\tikzset{BinaryFactor/.append style = {bend left}}
		\draw[BinaryFactor] (X0) to node[LabelLidar]{} (X3);
		\draw[BinaryFactor] (X4) to node[LabelLidar]{} (X2);
		
		\path (X4) edge[thick] (GPS)  ;
		\path (X0) edge[thick] (Prior)  ;
		
		\end{tikzpicture}
		\caption{MAP estimation as a factor graph: Blue circles denote the successive states of the vehicle $X_0,X_1,\cdots,X_N$ through times $0,1 \cdots ,N$. Factor nodes shown in black involve successive states and correspond to dynamical model relations \eqref{eq::prop}. Factors nodes shown in orange correspond to measurements involving a pair of state variables \eqref{eq::obs2}. The  factor nodes shown by a red circle and a blue diamond respectively represent an observation of the form \eqref{eq::obs}, and a prior on the initial state.}\label{fig:graph} 
	\end{figure}
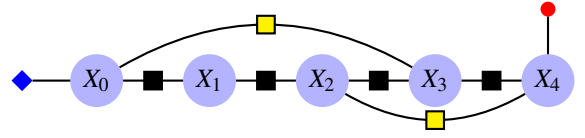
	A factor graph is a bipartite graph  that contains  \emph{variable nodes} $X_j \in\chi$ and \emph{factor nodes} $\eta_i$, each involving several variable nodes. There is an edge between a factor and each of the variables it involves. Factors may encode the model relation  \eqref{eq::prop} or   measurements of the form  \eqref{eq::obs} or  \eqref{eq::obs2}, and may be viewed as constraints between variables of the trajectory $\chi$ one seeks to estimate.  See Figure \ref{fig:graph} for a visualization.

	\subsection{Resolution of the nonlinear optimization problem}
	
	The minimization problem \eqref{eq::argmin} is commonly tackled resorting to Gauss-Newton or Levenberg-Mardquardt methods. The idea is to start from an initial guess of the trajectory, typically obtained through dead-reckoning, and successively \emph{linearize} the problem, and solve the obtained  linear least squares, that is, a quadratic  multidimensional optimization problem at each iteration.  Starting from an initial guess of the trajectory $\chi^{(0)}:=(X_0^{(0)},X_1^{(0)},\cdots,X_N^{(0)})$, the likelihood \eqref{eq::log-likelihood} may be linearized around $\chi^{(0)}$, and Gauss-Newton algorithms seek to compute
	\begin{equation}
	\label{eq::log-likelihood222}
	\delta\chi^* =   \argmin \bar L(\delta \chi),
	\end{equation}
	where  $\bar L$ is the linearized cost.  To keep the exposition simple, we assume in the sequel noises are additive, that is, \eqref{eq::prop} is of the form 
	$X_{k+1} = f(X_k,u_k) +w_k$, and also that $Z_k=h(\chi)+V_k$. Then, letting $\delta \chi:=\chi- \chi^{(0)}$, the linearized cost writes
	\begin{equation}
	\begin{aligned}
	\label{eq::log-likelihood22}
	\bar L(\delta \chi)&=||\delta X_0 - (\hat{X}_0 - X^{(0)}_0)||^2_{P_0} \\
	& + \sum_{k=0}^{N-1} || \delta X_{k+1}- F_k \delta X_k - (f(X^{(0)}_k, u_k) - X^{(0)}_{k+1}) ||^2_{Q_k} \\
	& + \sum_{k=1}^{N} || H_k \delta \chi - (Z_k - h_k(\chi^{(0)})) ||^2_{R_k}\end{aligned}
	\end{equation}
	where {  $F_k$ denotes  the Jacobian of $f$  with respect to $X_k$, and $H_k$ that of $h_k$ at current estimate $\chi^{(0)}$.
	} 
	After solving \eqref{eq::log-likelihood222}-\eqref{eq::log-likelihood22}, the linearization point is updated as $\chi^{(1)}=\chi^{(0)}+\delta\chi^*$ and serves as a new linearization point until convergence. 
	\begin{rem}
		This easily generalises for smoothing on-manifold, see for instance \cite{chauchat2018invariant, kuemmerle2011g2o, forster2016preintegration}, by simply changing the prior factor and update definitions, and adapting the jacobians accordingly. To save space, we omit it herein. 
	\end{rem}
	
	\subsection{Resolution of the linearized optimization problem}\label{sqrt:sec}
	
	We see at each step the   algorithm is faced with the resolution of the linearized optimization problem \eqref{eq::log-likelihood222}-\eqref{eq::log-likelihood22}. This is a standard least squares problem, and the solution comes in closed form.   Note indeed that  the log-likelihood $\bar L$ can be re-written as a single quadratic cost function:
	\begin{equation}
	\label{eq::quad}
	\bar L(\delta \chi) = ||A \delta \chi - b ||_{\Sigma}^2,
	\end{equation}
	with
	\begin{equation}
	\label{eq::quad22}
	A = \begin{pmatrix}
	\begin{matrix}
	I & 0 & ... & ... & 0 \\
	-F_1 & I & ... & ... & 0 \\
	... & ... & ... & -F_N & I \\
	\end{matrix} \\
	H_1 \\
	\vdots \\
	H_N
	\end{pmatrix},
	b =
	\begin{pmatrix}
	a_0 \\
	a_1 \\
	\vdots \\
	a_N \\
	c_1 \\
	\vdots \\
	c_N
	\end{pmatrix}
	\end{equation}
	where $\Sigma = \text{blkdiag}(P_0,Q,R)$, where $Q$ and $R$ are block diagonal matrices stacking the sequences of covariance matrices $Q_n$ and $R_k$, $a_0 = \hat{X}_0 - X^{(p)}_0$, $a_i = (f(X^{(p)}_{i-1}, u_{i-1}) - X^{(p)}_{i})$ if $i>0$, and $c_k = (Z_k - h_k(\chi^{(p)}))$. Computing the gradient of \eqref{eq::quad} and setting it to zero, allows one to conclude the  minimizer $\delta \chi^*$ satisfies the \emph{normal equations}:
	\begin{equation}
	\label{eq::inverse}
	(A^T \Sigma^{-1} A) \delta \chi^* = A^T \Sigma^{-1}b
	\end{equation}
	Solving the normal equations \eqref{eq::inverse} is the main task to be addressed when implementing factor  graph optimization.  Indeed, one is faced with a system of linear equations. Most popular methods rely on   factorizations of the information matrix defined by  $ \mathcal I:=A^T \Sigma^{-1} A$ in the form of $A^T \Sigma^{-1} A=LL^T$ with $L$ lower triangular  (Cholesky) or through a QR factorization. Matrix $L$ is referred as the ``square root'' of the information matrix, and allows to solve the problem by first solving $L\eta=A^T \Sigma^{-1}b$ and then $\eta=L^T \delta \chi^* $ by back-substitution, see \cite{dellaert2006square}.

	\subsection{Potential issue n$^\circ$1: computational complexity }
	
	$A$ is a   matrix having    size $((N+1)d_x+Kd_z)\times ((N+1)d_x)$ where $d_x,d_z$ are the dimensions of the state and observation vectors,  $K$ is the number of observations (we assumpe for simplicity all observations have identical dimension).  In principle,  the problem would not be tractable for localization applications. However, the sparsity of $A$ stemming from the particularity of the problem estimation structure allows smoothing solvers to encode various ways of inverting this equation. Indeed, each factor only involves few variables (usually one or two) \cite{eustice2006exactly}. For instance, \cite{kaess2012iSAM2} embeds the problem in a particularly well adapted Bayes tree structure which  allows for local constant-time  updates at each time step by reusing previous factorizations. In a more recent work, \cite{liu2018ice-ba} uses preconditioned conjugate gradient to invert the problem, with a high efficiency gain coming from a number of computational tricks specific to the SLAM problem. These methods usually prefer working with an equivalent formulation of \eqref{eq::quad}:
	$$
	\bar L(\delta X) = || \tilde{A} \delta \chi - \tilde{b} ||^2
	$$
	with $\tilde{A} = \Sigma^{-1/2} A$ and $\tilde{b} = \Sigma^{-1/2} b$, where $\Sigma^{-1/2}$ is the square-root of $\Sigma^{-1}$. However, both formulations explicitly rely on the inverse of the covariance matrices at play.
	
	\subsection{Potential issue n$^\circ$2: ill-conditioned normal equation matrix}The complexity issue is obvious, and much efforts  in the SLAM back-end community have been devoted to it. Another potential issue that has scarcely been addressed so far, and that we bring forward in the present paper, is as follows.  If the normalization matrix  $\Sigma^{-1}$ is ill conditioned because $\Sigma$ has very small eigenvalues (or even null eigenvalues), then measurement noise is amplified and the solution may become grossly inaccurate.  In the present paper, we focus on the case where dynamics \eqref{eq::prop} rely on high precision inertial sensors, which are becoming increasingly common, and lead to very small covariance matrices.  
	
	Assume  the covariance  matrix $Q$  has very small eigenvalues. As $Q\to 0$, we see that $\Sigma^{-1}\to\infty$ and thus 
	\begin{equation*}
	\begin{matrix}
	(A^T \Sigma^{-1} A)  \longrightarrow \infty, & 
	(A^T \Sigma^{-1} b)  \longrightarrow \infty ,\\
	\Sigma^{-1/2} A  \longrightarrow \infty, & 
	\Sigma^{-1/2} b \longrightarrow \infty.
	\end{matrix}
	\end{equation*}
	But this does not mean that the quantity of interest $\delta \chi^*$ degenerates in  the same way, as we have an ``$\frac{\infty}{\infty}$'' indetermination from \eqref{eq::inverse}. In fact, it does converge to a finite value $\delta \chi_{\infty}$
	which is simply the result of a lower-dimensional problem:  we will prove in the next section that when $Q\to 0$ we have 
	\begin{equation}
	(A^T \Sigma^{-1} A)^{-1}A^T \Sigma^{-1} b \longrightarrow \delta \chi_{\infty},
	\label{eq::limit}
	\end{equation}Ideally, we would like the solver to be such that $\delta \chi^*\to \delta \chi_{\infty}$ when $Q\to 0$, and also to be able to find $ \delta \chi_{\infty}$ even in the degenerate case where $Q=0$. However, we cannot expect such a desirable behavior from a solver directly based on the normal equations \eqref{eq::inverse}. This is why we advocate in the present paper a different approach based on the Kalman smoother to solve the linearized optimization problem \eqref{eq::log-likelihood222}, \eqref{eq::log-likelihood22}.  
	
	\subsection{A  solution without matrix inversion: the  robust batch solver}\label{sec:batch_solver}
	
	Assume   the noise matrix of the model $Q$ should not be inverted. Owing to the block-diagonal structure of the noise matrix $\Sigma$, the linearized cost  \eqref{eq::quad} may be split into two parts as follows:
	\begin{equation}
	\bar L({\delta \chi}) =\|A_1 \delta \chi - b_1 \|_{\Sigma_1}^2 + \|A_2 \delta \chi - b_2 \|_{\Sigma_2}^2
	\label{eq:separate_lq}
	\end{equation}
	that is, $A = \begin{pmatrix} A_1 \\ A_2 \end{pmatrix}$, $b = \begin{pmatrix} b_1 \\ b_2 \end{pmatrix}$, and $\Sigma = \begin{pmatrix} \Sigma_1 & 0\\0 & \Sigma_2 \end{pmatrix}$, and where  $\Sigma_1$ should not be inverted, and where we require $A_1$ to be square and invertible. In the meantime, $\Sigma_2$ should be well-conditionned. There might be less factors with close to singular covariances than the dimension of the state, and $A_1$ may be completed with additional factors having well-conditioned covariances to make it square if need be.
	Using the associated normal equations, and the matrix inversion lemma, the minimizer of the cost \eqref{eq:separate_lq} reads:
	\begin{equation}
	\begin{aligned}
	\delta \chi^* = A_1^{-1} ((I - K J) b_1 + K b_2), \\
	K = \Sigma_1 J^T (J \Sigma_1 J^T + \Sigma_2)^{-1}, \quad J = A_2 A_1^{-1}.
	\label{eq:robust_solution_batch}
	\end{aligned}\end{equation}
	The inverted term is the sum of a possibly ill-conditionned matrix with small eigenvalues, and a well-conditionned one, and should thus have no eigenvalue close to zero. 
	Wee are  now in a position to prove the result announced in equation \eqref{eq::limit}. 
	\begin{lem}
		As  $Q\to 0$, the solution $\delta\chi^*$  to \eqref{eq::inverse} tends to a finite value.
	\end{lem}
	\begin{IEEEproof}
		Expanding the definitions of $K$ and $J$  in  \eqref{eq:robust_solution_batch}, we obtain:
		$ 
		\delta \chi = A_1^{-1} \left[ b_1 + \Sigma_1 A_1^{-1}A_2^T (A_2 A_1^{-1}\Sigma_1 A_1^{-T}A_2^T+\Sigma_2)^{-1} (b_2 - A_2 A_1^{-1} b_1) \right]
		$. 
		We see $(A_2 A_1^{-1}\Sigma_1 A_1^{-T}A_2^T+\Sigma_2)\to (A_2 A_1^{-1}\Sigma_\infty A_1^{-T}A_2^T+\Sigma_2)$ when $\Sigma_1\to 0$, and this latter term   is lower-bounded by $\Sigma_2$, so there is no indetermination   in \eqref{eq::limit}: $\delta X_{\infty}$ exists.
	\end{IEEEproof}
	
	This solution may become intractable as it requires the full inverse of $A_1$, and the size of the matrix $A_1$ is quadratic in the size of the trajectory.  According to \eqref{eq::quad22},  $A_1$ has a  lower block-triangular structure, and its inverse may be obtained analytically, but  the number of linear systems to solve because of the term $J = A_2 A_1^{-1}$ still is a major caveat of this method. It is however used in Section \ref{sec:data_joint_lab} as a reference for comparison purposes. In the next sections we investigate an approach based on Kalman smoothers, which reaches identical result, but with a  complexity being linear in the length of the trajectory. 
	
	\section{The linear Kalman smoother as a least squares solver}\label{BIFM:secc}
	
	Consider the problem of finding the minimum of cost function $\bar L$ given by \eqref{eq::log-likelihood22}.  In this section we will recall the Kalman smoother may serve as a solver to this optimization problem, albeit (for now)  in the particular case of unary observations of the form \eqref{eq::obs}.
	
	\subsection{The standard Kalman smoother}
	Consider the following linear system with unary observations:
	\begin{equation}\begin{aligned}\label{Kalman1}X_{k+1}&=F_k X_k + u_k + w_k,\\ Z_k&=H_kX_k+V_k\quad\text{(unary observations)}\end{aligned}\end{equation} $Z_0 = X_0 + V_0$ encodes the prior on $X_0$.  It is  easily proved that  $\bar L$ given by  \eqref{eq::log-likelihood22}  is related to this linear system as   
	$$\bar L=-\log p(X_0,\ldots,X_N\mid Z_1,\ldots, Z_N)+c,$$
	where $c \in \RR$ is a constant.
	The linear Kalman smoother  computes the MAP estimate which minimizes $\bar L$, and thus may output the quantity of interest $\delta\chi^*$ that is required at each linearization step of the more general Gauss-Newton solver. Its implementation is generally based on a forward recursion (e.g. the Kalman filter), and a backward recursion which backpropagate the information to the past states, for instance the standard Rauch-Tung-Striebel (RTS) implementation \cite{rauch1965maximum}.  
	
	\subsection{The Backward Information Filter Forward Marginal (BIFM)}\label{BIFM:sec}
	
	The RTS implementation of the Kalman Smoother avoids inverting the propagation covariance, but not the states' ones, which appear in the gain used for the backpropagation \cite{rauch1965maximum}. In this work, we propose to use an alternative formulation of the Kalman smoother equations, the Backward Information Filter Forward Marginal (BIFM), which specifically avoids inverting the states' covariances \cite{loeliger2016sparsity}. This approach has already been used for message passing in Gaussian factor graphs in signal processing, but only for acyclic graphs. Another version of the Kalman smoother, namely the Modified-Bryson Frasier smoother, also exists \cite{bierman1977factorization}. Square-root forms of both algorithms have been produced and were compared on various signal processing problems in  \cite{wadehn2016square}. BIFM appeared slightly superior in these cases. Moreover, it can be derived in a very ``Kalman filter like'' manner which does not require writing the inverse of the covariance matrices $Q_k$ nor the inverse of the forward covariance matrix $P_k$ defined in \eqref{kalak}, as we are about to prove.

	Consider the system \eqref{Kalman1}. {BIFM is based on the following approach. Suppose   $P(X_k | Z_0, \cdots, Z_k)\sim\mathcal N(x_k^f,P_k^f)$; and likewise $P(X_k | Z_{k+1}, \cdots, Z_N)\sim\mathcal N(x_k^b,P_k^b)$. Here the superscripts stand for forward and backward. $x_k^f$ is obtained by a Kalman filter. Then $x_k^b$ may be considered as a measurement of $X_k$ with noise covariance $P_k^b$, and treated as such in a Kalman update to merge it with $(x_k^f, P_k^f)$ and obtain a final estimate $\hat{X}_k$. As there is no prior for the backward phase, that is, the prior on $x_0^b$ is ``flat'', the information form of the Kalman filter must be used \cite{maybeck1982stochastic}, starting with zero information prior. The detailed equations are given in the following. For the sake of readability, the forward and backward distributions are denoted by, for all $k$ :
		\begin{align}
		P(X_k|Z_{0},\dots,Z_k) &\sim \mathcal{N}(x_{k},P_{k}) \label{kalak}\\
		P(X_k|Z_{k+1},\dots,Z_N) &\sim \mathcal{N}(J_k^{-1}y_k,J_k^{-1}),
		\end{align}
		where $y_k$ and $J_k$ are the information vector and matrix respectively.
	} 
\begin{rem}
	Note that, here, the information matrix $J_k$ is the inverse of a covariance, while the information matrix referred to in Section~\ref{sqrt:sec}, $\mathcal I$, is the Fisher information matrix.
\end{rem}
	The forward recursion may be performed based on the standard Kalman filtering belief. The parameters of the Gaussian are returned by the standard linear Kalman filter by alternating between the propagation step:\begin{equation}
	\begin{aligned}
	x_{k+1|k} &= F_k x_{k}+u_k\\
	P_{k+1|k} &= F_k P_{k} F_k^T + Q_k
	\end{aligned}\label{prop:::eq}\end{equation}
	and letting $K_{k+1}:=P_{k+1|k} H_k^T (H_k P_{k+1|k} H_k^T + R_k)^{-1}$, the update step\begin{equation}
	\begin{aligned}
	x_{k+1} &=  x_{k+1|k} + K_{k+1} (Z_k - H_k x_{k+1|k})
	\\
	P_{k+1} &= P_{k+1|k} - K_{k+1} H_k P_{k+1|k}.
	\end{aligned}\label{up:::eq}\end{equation} 
	We see that even if some or even all the eigenvalues of  $P_0$ and $Q$  are null, the computation may be performed as long as $R$ is non-singular. In particular, this means that the measurement noise must be sufficiently large for BIFM to be applied. As a consequence, $Q$ can be as small as desired without leading to numerical issues.
	{The equations for $J_k$ and $y_k$ in the backward recursion are obtained by setting $J_N$ to be zero, and then considering \eqref{Kalman1} in reverse time, and combining it with \eqref{prop:::eq} and \eqref{up:::eq}, using the identities $x_k = J_k^{-1} y_k$, $P_k = J_k^{-1}$. The update in information form simply becomes:
		\begin{equation}
		\label{eq:backward_upd}
		\begin{gathered}
		J_{k|k} = J_k + H_k^T R_k^{-1} H_k, \\
		y_{k|k} = y_k + H_k^T R_k^{-1} Z_k
		\end{gathered}
		\end{equation}
		The backpropagation equations read:
		\begin{equation}
		\label{eq:backward_prop}
		\begin{gathered}
		J_k = F_k^T (I + J_{k+1|k+1} Q_k)^{-1} J_{k+1|k+1} F_k, \\
		y_k = F_k^T (I + J_{k+1|k+1} Q_k)^{-1} (y_{k+1|k+1} - J_{k+1|k+1} u_k)
		\end{gathered}
		\end{equation}

		And the final solution $X_k^* = \argmax_{X_k}P(X_k|Z)$ is given by merging the estimates obtained at the forward and backward pass as follows for each $k$, see \cite{loeliger2016sparsity}:
		\begin{equation}
		X_k^* = x_k + P_k (P_k +J_k^{-1})^{-1} (J_k^{-1} y_k - x_k).
		\label{eq:final_upd_init}
		\end{equation}
		However, the given formula implies inverting $J_k$, which we want to avoid. Therefore, we propose to modify \eqref{eq:final_upd_init} as follows:
		\begin{align}
		X_k^* &= x_k + P_k (P_k +J_k^{-1})^{-1} J_k^{-1} (y_k - J_k x_k) \\
			&= x_k + (I + P_k J_k)^{-1} (P_k y_k - P_k J_k x_k) \\
			&= (I + P_k J_k)^{-1} (x_k + P_k y_k),
		\label{eq:final_upd}
		\end{align}
		where the push-through identity \cite{henderson1981deriving} shows that $P_k (P_k +J_k^{-1})^{-1} J_k^{-1} = (I + P_k J_k)^{-1} P_k$. Moreover, the final covariance of $X_k^*$ given $Z$, $P_{k|N}$ is given by
		\begin{equation}
		P_{k|N} = (I + P_k J_k)^{-1} P_k
		\label{eq:final_cov}
		\end{equation}
	} 
	We see the  equations above allow performing  optimal smoothing without involving at any time the matrix inverses $Q^{-1}$ or $P_k^{-1}$ where $P_k$ denotes the forward covariance matrices, see \eqref{kalak}.  Indeed, in \eqref{eq:backward_prop} and \eqref{eq:final_upd_init} each time those matrices are involved in an inversion operation there is a natural regularization term $(I+\cdot)$ involved as well.

	\subsection{Summary of the approach}
	
	We summarize the approach and the results obtained so far.  
	\begin{itemize}
		\item We consider nonlinear system \eqref{eq::prop} with measurements of the form \eqref{eq::obs} and \eqref{eq::obs2}. In the smoothing approach, one seeks to compute the most likely (entire) trajectory $\chi$ of  \eqref{eq::traj} in the light of  past measurements. By contrast,  the filtering approach is only concerned with computation of the most likely current state $X_N$. 
		\item The solution to the smoothing problem corresponds to optimization problem  \eqref{eq::argmin} with cost function \eqref{eq::log-likelihood}. To attack this problem one usually linearizes the cost at current estimate and solves a simplified optimization problem \eqref{eq::log-likelihood222} with quadratic cost function \eqref{eq::log-likelihood22}, that is, a least squares problem. This provides a correction to the current estimate of $\chi$, and then the cost  \eqref{eq::log-likelihood} is relinearized at the corrected estimate, yielding another least squares problem which in turn provides a new correction. This is repeated until convergence to the optimum. 
		\item At each optimization step, the least squares solution \eqref{eq::log-likelihood222} involves solving the normal equations \eqref{eq::inverse}, which may be ill-conditioned when process noise is too low. In the present paper, we address the problem of smoothing in the presence of process noise covariance matrices that may be ill-conditioned or even singular,  owing to the use of highly accurate motion sensors.
		\item In the case of unary observations \eqref{eq::obs} only, one may associate a linear dynamical system \eqref{Kalman1} with the linearized cost \eqref{eq::log-likelihood22} at each step. The Kalman smoother then provides a solution to the optimization problem \eqref{eq::log-likelihood222}. By using a slightly modified implementation related to the BIFM of \cite{loeliger2016sparsity}, we obtain a solution to \eqref{eq::log-likelihood222} without inverting matrices that may be ill-conditioned or singular. 
		\item In the sequel, we seek to adapt the proposed latter solution to the case where measurements of the form \eqref{eq::obs2} are also involved, as typically arises when using vision, notably for SLAM. 
	\end{itemize}

	\section{Proposed method: the BIFM Kalman smoother with stochastic cloning}
	\label{sec:new_solver}
	Throughout this section, we still consider the linearized optimization problem \eqref{eq::log-likelihood222} with   cost   \eqref{eq::log-likelihood22}, but where observations may involve pairs of states, to account for measurements of the form \eqref{eq::obs2}. As previously done in \eqref{Kalman1}, one may then associate   a linear system
	\begin{align}X_{k+1}&=F_k X_k + u_k + w_k,\label{Kalman12}\\ Z_k&=H_kX_k+V_k\quad\text{(unary observations)}\label{Kalman13}\\Z_k &= H_k X_k + H_l X_l + V_k,~l<k ~\text{(pair of states observations)}\label{Kalman14}\end{align}
	to the optimization problem, where \eqref{Kalman14} stems from the linearization of observations of the form \eqref{eq::obs2}. As in Section~\ref{sec:observations}, $Z_k$ denotes both types of measurements. Building upon the BIFM of the previous section, we will attempt to solve the corresponding optimization problem (i.e., smoothing) without ever inverting  matrices that may be ill-conditioned or singular in the presence of low or even null process noise $w_k$.

	\subsection{Stochastic cloning for filtering}
	\label{sect::cloning}
	
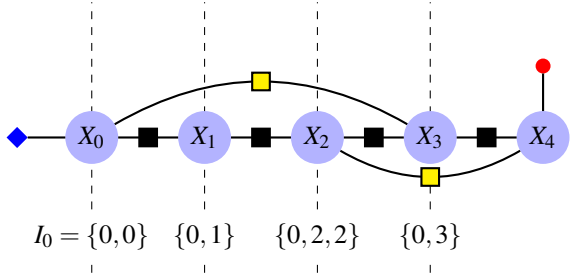
\begin{figure}
		\centering
		\begin{tikzpicture}[node distance = 1.5cm]
		\tikzset{Variable/.style = {shape          = circle,
				fill           = blue!30,
				text           = black,
				inner sep      = 2pt,
				outer sep      = 0pt,
				minimum size   = 20 pt}}
		\tikzset{BinaryFactor/.style   = {thick,
		}}
		\tikzset{LabelIMU/.style =   {draw,
				fill           = black}}
		\tikzset{LabelLidar/.style =   {draw,
				fill           = yellow}}
		\tikzset{GPS/.style = {shape          = circle,
				fill           = red,
				text           = black,
				inner sep      = 2pt,
				outer sep      = 0pt,
				minimum size   = 5 pt}}
		\tikzset{Prior/.style = {shape          = diamond,
				fill           = blue,
				text           = black,
				inner sep      = 2pt,
				outer sep      = 0pt,
				minimum size   = 5 pt}}
		
		\path [draw] [dashed] ($(X0)+(0,-1.8)$) -- node[above, pos=0.05, fill=white] {$I_0 = \{0,0\}$} ($(X0)+(0,1.8)$);
		\path [draw] [dashed] ($(X1)+(0,-1.8)$) -- node[above, pos=0.05, fill=white] {$\{0,1\}$}($(X1)+(0,1.8)$);
		\path [draw] [dashed] ($(X2)+(0,-1.8)$) -- node[above, pos=0.05, fill=white] {$\{0,2,2\}$} ($(X2)+(0,1.8)$);
		\path [draw] [dashed] ($(X3)+(0,-1.8)$) -- node[above, pos=0.05, fill=white] {$\{0,3\}$} ($(X3)+(0,1.8)$);
		
		\node (X0) [Variable] {$X_0$};
		\node (X1) [Variable,right of=X0] {$X_1$};
		\node (X2) [Variable,right of=X1] {$X_2$};
		\node (X3) [Variable,right of=X2] {$X_3$};
		\node (X4) [Variable,right of=X3] {$X_4$};
		
		\node (GPS) [GPS, above = 0.5cm of X4] {};
		\node (Prior) [Prior, left = 0.5cm of X0] {};
		
		\draw[BinaryFactor] (X0) to node[LabelIMU]{} (X1);
		\draw[BinaryFactor] (X1) to node[LabelIMU]{} (X2);
		\draw[BinaryFactor] (X2) to node[LabelIMU]{} (X3);
		\draw[BinaryFactor] (X3) to node[LabelIMU]{} (X4);
		
		\tikzset{BinaryFactor/.append style = {bend left}}
		\draw[BinaryFactor] (X0) to node[LabelLidar]{} (X3);
		\draw[BinaryFactor] (X4) to node[LabelLidar]{} (X2);
		
		\path (X4) edge[thick] (GPS)  ;
		\path (X0) edge[thick] (Prior)  ;
		\end{tikzpicture}
		\caption{Stochastic cloning methodology for the forward pass in the factor graph framework. At each time $k$, state variable $X_k$ is augmented to form the variable $\tilde X_k:= X_{I_k}$ where $I_k$ consists of current state index $k$ and all  the indices of the current and past variables  that are to be used later in a relative measurement. In a factor graph representation, variables involved in stochastic cloning are easily visualized:  ``clones'' at time $k$ are the variables related by an orange factor that spans from, to, or ``over'' the current state.}\label{fig:graph2} 
	\end{figure}

	Although we are concerned with smoothing, we have seen the Kalman filter is the main component of  the forward pass of the BIFM smoother. 
	Standard Kalman filtering for linear systems of the form \eqref{Kalman12}-\eqref{Kalman14} has been rendered possible through the stochastic cloning method, introduced in  \cite{mourikis2007sc-kf}  and one of the key components of the MSCKF \cite{mourikis2007multistate}.  The state is cloned at some point in time $l$  and kept in memory to be able to compute an update that will involve it in the future at time $k$, see \eqref{Kalman14}. The clones are discarded once they are not useful anymore. Note that  the variables designated as clones are  easily visualized in the factor graph framework, see Figure \ref{fig:graph2}.
	\begin{rem}
		Since the focus is put on the linearized problem, the factor graph is fixed, therefore we know in advance which states are linked together, and thus when cloning or discarding. Removing a clone does not delete it from the higher-level estimation process, as does marginalisation for instance.
	\end{rem}
	To incorporate a measurement \eqref{Kalman14} in the Kalman filter update, one needs  the mean and covariance matrix of the vector $(X_l,X_k)$. Thus, denoting by $I_k$ the set containing $k$ and all indices of past states involved in future measurements, we may define the current state at time step $k$ as $\tilde X_k=(X_i)_{i \in I_k}$, and we see its size changes over time, see Figure \ref{fig:graph2}.  
	
	\subsubsection{The stochastic cloning pipeline}
	
	The goal of Kalman filtering with SC is to compute at each time step $k$ the parameters of the Gaussian density $P((X_i)_{i \in I_k}|Z_0,\dots Z_{k-1})$, and  then to update them accounting for observations available at $k$, that is, compute the parameters of  distribution $P((X_i)_{i \in I_k}|Z_0,\dots Z_k)$. This can be done along the lines of the standard Kalman filter, even though the  dimension of the state keeps changing over time. 
	The cloning pipeline may be intuitively described  as follows:
	$$\text{Cloning} \xrightarrow{} \text{Propagation} \xrightarrow{} \text{Update} \xrightarrow{} \text{Clone discarding}$$
	Let $\tilde{P}_k$ denote the covariance matrix of the augmented state $\tilde{X}_k = (X_i)_{i \in I_k}$. The SC steps are as follows. 
	\paragraph*{Clone creation (duplication step)}The cloning step corresponds to the fact that if the current state $X_k$ is to be used in a future relative measurement, then it must  be cloned. This simply consists in duplicating the $X_k$ and concatenating it with the full state $\tilde{X}_k$, which  thus  now contains the state $X_k$ twice. Indeed, the current state is of the form $\tilde{X}_k$ where $I_k$ contains current index $k$ and index of  clones already created and not yet discarded. Duplication of current state then writes:
	\begin{equation}
	\tilde{X}_k \leftarrow( X_k,\tilde{X}_k)
	\label{eq:cov_clone2}
	\end{equation}
	This may be rewritten in terms of matrix computation as follows:
	\begin{equation}
	\tilde{X}_k \leftarrow C_k \tilde{X}_k,\qquad C_k =\begin{pmatrix}
	I & &\\
	I & &\\
	& \ddots & \\
	& & I
	\end{pmatrix}\label{eq:cloning_matrix}
	\end{equation}
	Since $X_k$ and its copy are fully correlated, the covariance of the extended vector must be replaced with
	\begin{equation}
	\tilde{P}_k \leftarrow C_k \tilde{P}_k(C_k)^T= \begin{pmatrix} P_k & P_{{I_k},k} \\ P_{k ,{I_k}} & \tilde{P}_k \end{pmatrix},
	\label{eq:cov_clone}
	\end{equation}
	where $P_k$ is the marginal covariance of  the current state variable $X_k$,   $\tilde{P}_k$ the marginal covariance of the augmented state $\tilde{X}_k$ before duplication,  $P_{I_k,k}$  and  $P_{k , {I_k}}$ the cross-correlations.  The computation of $P_{k , {I_k}}$ is made by copying blocks from $\tilde{P}_k$  and  the correlation between $X_k$ and $X_k$ is the identity matrix.
	
	\paragraph*{Propagation}
	Clones  remain static during propagation, since the rationale is merely to keep past states in the augmented state. Hence we use a propagation step analogous to \eqref{prop:::eq} but with   augmented dynamics, input, and process noise as follows:
	\begin{equation}
	F_k^p = \begin{pmatrix}
	F_k & 0\\
	0 & I
	\end{pmatrix} ,\quad
	\tilde{u}_k = \begin{pmatrix}
	u_k \\
	0 
	\end{pmatrix}, \quad
	\tilde{w}_k = \begin{pmatrix}
	G_k w_k \\
	0 
	\end{pmatrix}.
	\label{eq:clone_propagation}
	\end{equation}
	
	\paragraph*{Update}
	The main purpose of SC is to allow carrying out the updates as in the standard Kalman filter. Indeed, as each relative observation at time $k$ involves a past state variable with index $l<k$ that has been cloned, and thus $l,k\in I_k$,  observations \eqref{Kalman14} may be written using a  novel observation matrix  $
	\tilde{H}_k = (H_i)_{i \in I_k},
	$ defined such that 
	\begin{equation}
	\tilde{H}_kX_{I_k}= \sum_{i \in I_k} H_i X_i.
	\label{eq:clone_updatee}
	\end{equation}

	\paragraph*{Clone discarding}
	Once a clone is no longer useful (i.e., it will    be involved in no later measurement), it can be discarded. This is done by marginalizing out the considered clone,   which reads in terms of matrices
	\begin{equation}
	\tilde{X}_k \leftarrow D_k  \tilde{X}_k,  \quad \tilde{P}_k \leftarrow D_k \tilde{P}_k (D_k)^T,\label{discarme:eq}
	\end{equation}where $D_k $ is the identity matrix from which we removed the rows corresponding to states we  discard, that is, those in $I_k \setminus I_{k+1}$. 
	
	\subsubsection{SC implementation}
	
	Gathering all the steps above, we see the obtained Kalman filter maintains the augmented state  $\tilde{X}_k = (X_i)_{i \in I_k}$ and its covariance matrix $\tilde{P}_k$. Formally, $\tilde{X}_k$ follows the state equations
	\begin{gather}
	\label{eq::augmented_prop}
	\tilde{X}_{k+1} = \tilde{F}_k \tilde X_{k} + \tilde{u}_k + \tilde{w}_k, \\
	Z_k = \tilde{H}_k \tilde X_{k} + V_k,
	\label{eq::augmented_obs}
	\end{gather}
	using the augmented quantities defined in \eqref{eq:clone_propagation}, and where the  matrix $\tilde{F}_k$  is defined with the help of matrices introduced in  \eqref{eq:cloning_matrix}, \eqref{eq:clone_propagation}, and \eqref{discarme:eq},  by
	\begin{equation}
	\tilde{F}_k = F_k^p C_k D_k,
	\label{Fd:eq}
	\end{equation}
	(note we suggest to perform the discarding step before duplication to spare the algorithm undesirable computations), and with observations defined by  \eqref{eq:clone_updatee} corresponding to \eqref{Kalman13}-\eqref{Kalman14}. Formally the obtained system \eqref{eq::augmented_prop}-\eqref{eq::augmented_obs} fits into the standard system form used by the Kalman filter, and in fact the dimension of the state changes does not result in any modification of the Kalman filter method.

	\subsection{Proposed ``SC-BIFM''  (stochastic cloning for smoothing)}
	\label{sect::solver}
	
	The proposed solver consists in applying the BIFM smoother of Section \ref{BIFM:sec} dedicated to system \eqref{Kalman1}  to the augmented system created by stochastic cloning to address systems of  the form \eqref{Kalman12}-\eqref{Kalman13}-\eqref{Kalman14}. This way, we will achieve robustness  to   singular prior and process noise covariance matrices.

	\subsubsection{Forward pass with stochastic cloning}
	The forward pass is akin to Kalman filtering, and the method of  Kalman filtering with stochastic cloning was recalled in \eqref{prop:::eq}-\eqref{up:::eq} and Section~\ref{sect::cloning}, and put in perspective in the proposed factor graph context. The forward pass consists of the BIFM forward recursion \eqref{prop:::eq}-\eqref{up:::eq}, applied to the augmented system  \eqref{eq::augmented_prop}-\eqref{eq::augmented_obs} having augmented state variables $\tilde X_k$. 
	
	\subsubsection{Backward stochastic cloning with an information filter} 
	This step consists in applying the backward equations in information form  \eqref{eq:backward_prop} to the augmented system   \eqref{eq::augmented_prop}-\eqref{eq::augmented_obs} instead of the standard system dynamics with unary observations \eqref{Kalman1}.
	As a noticeable difference, we observe that contrary to $F_k$ in \eqref{Kalman1}, matrix  $\tilde{F}_k$ is not  square. However,   only its transpose is involved in the backwards equations and this poses no problem.  This is an advantage of having written the backwards recursion in information form: in standard form writing state at time $k$ from state at time $
	k+1$   would require matrix inversion.

	Let $\tilde y_k$ be the augmented backward information vector, and $\tilde J_k$ the associated information matrix. Let us comment on the steps involved in the backwards pass in the presence of clones.
	
	\paragraph*{Clone creation}
	In the forward  pass, the vector state is augmented each time a pose has to be kept for later use, while the covariance matrix is augmented with full correlation as in Eq. \eqref{eq:cov_clone}. In the backward case, the order of the steps is reversed, so the information vector and information matrix are augmented each time a clone was discarded in the forward  pass. The process of clone creation and destruction in backwards time strictly  mirrors the forward time. However when  augmentation occurs in the backward sense  new entries of both $\tilde y_k$ and $\tilde J_k$ are padded with zeros: we create a state with no correlation with the rest of the system. Note indeed that the clone refers to another state variable $X_k$ appearing in the ``past'', that is, $X_l$ with $l<k$. Thus immediately after creating it the observation linking the current state $k$ to the added state having index $l$ is performed, which will create correlation. Note this interesting fact: in the backward sense the clone does not ``know'' it is a clone (i.e., is by no means \emph{fully} correlated with any state but only loosely correlated to it via the noisy  observations involving the pair of states). Full correlation only appears when it gets discarded, mirroring the forward case where the clone is ``informed'' it is a clone, through Eq. \eqref{eq:cov_clone}, at its creation.

	\paragraph*{Clone discarding}
	In the backward propagation, the counterpart of clone creation is surprisingly simple, since it boils down to applying $(D_k)^T$, see \eqref{Fd:eq} and recall from \eqref{discarme:eq}
	that  $D_k $ is the identity matrix from which we removed the rows corresponding to states we can discard.  Assuming the clone lies at position $l$ in the information vector and we are dealing with  state  at position $k$, the corresponding operation reads:
	$$
	\tilde y_k \leftarrow \tilde y_k + \tilde y_l
	$$
	\begin{equation}
	\label{eq:uncloning}
	\tilde J_{k,k} \leftarrow \tilde J_{k,k} + \tilde J_{k,l} + \tilde J_{l,k} + \tilde J_{l,l}
	\end{equation}
	
	{ These formulas are directly derived from  \eqref{eq:backward_prop} and \eqref{eq:cloning_matrix}  reading what happens to the entries of the matrix $J$ and the vector $\tilde y$. Indeed, the transpose of $C_k$ is applied, which sums the last two blocks of a vector. Combined with the transpose of the clone discarding matrix $D_k$, we get the above formulas. 
	} 
	
	\subsubsection{Interpretation}Although the backward pass was obtained through ``blind'' matrix manipulations of the system in reverse time, the following interesting interpretation of what cloning means in reverse time provides insight. Indeed,  discarding the clone in backward time  is actually a sequence of two actions: informing the clone it is a clone, then killing it (mirroring clone creation and full correlation information \eqref{eq:cov_clone} in the forward pass). The first action is equivalent to a noise-free observation $Z_k=X_k-X_l$, taking value $Z_k=0$. 
	{  To study this operation mathematically, let us associate a non-zero covariance matrix $R$ to observation $Z_k$. We will then study what happens when $R\to 0$. Since we are in information form, only the components $\tilde y_k,\tilde y_l$ and the blocks $\tilde J_{k,k}, \tilde J_{k,l}, \tilde J_{l,k}, \tilde J_{l,l}$ are affected by this operation. The Kalman update equations then yield:
	} 
	\begin{equation}
	\label{eq::obs_step}
	\begin{aligned}
	\tilde{J}^+ = 
	\begin{pmatrix}
	\tilde J_{k,k}^+ & \tilde J_{k,l}^+ \\
	\tilde J_{l,k}^+ & \tilde J_{l,l}^+
	\end{pmatrix}
	& = 
	\begin{pmatrix}
	\tilde J_{k,k} & \tilde J_{k,l} \\
	\tilde J_{l,k} & \tilde J_{l,l}
	\end{pmatrix}
	 +
	\begin{pmatrix} I & -I \end{pmatrix}^T
	R^{-1}
	\begin{pmatrix} I & -I \end{pmatrix} \\\
	& = \begin{pmatrix}
	\tilde J_{k,k} +R^{-1} & \tilde J_{k,l} -R^{-1} \\
	\tilde J_{l,k} -R^{-1} & \tilde J_{l,l} +R^{-1}
	\end{pmatrix}
	\end{aligned}
	\end{equation}
	After this observation, we marginalize out the clone. In information form, the remaining block $\tilde J_{k,k}^{++}$ is given by the classical Schur complement formula:
	$$
	\tilde J_{k,k}^{++} \leftarrow \tilde J_{k,k}^+ - \tilde J_{k,l}^+ \left[\tilde J_{l,l}^+\right]^{-1} \tilde J_{l,k}^+
	$$
	Replacing the blocks of $\tilde{J}^+$ by their values from Eq. \eqref{eq::obs_step}, the new block $\tilde J_{k,k}^+$ after the sequence `` observation + marginalization'' reads:
	\begin{align}
	\tilde J_{k,k}^{++} & = \tilde J_{k,k} + R^{-1} - \left( \tilde J_{k,l} - R^{-1} \right) \left( \tilde J_{l,l} + R^{-1} \right)^{-1} \left( \tilde J_{l,k} - R^{-1} \right) \\
	& = \tilde J_{k,k} + R^{-1} - \left( \tilde J_{k,l} R - I \right) \left( \tilde J_{l,l} R + I \right)^{-1} \left( \tilde J_{l,k} - R^{-1} \right) \\
	& = \tilde J_{k,k} + R^{-1} - \left( \tilde J_{k,l} R - I \right) \left( I - \tilde J_{l,l} R + O(R^2) \right) \left( \tilde J_{l,k} - R^{-1} \right),
	\end{align}
	{where, after factorising $R^{-1}$ in $\left( \tilde J_{ll} + R^{-1} \right)^{-1}$, the first-order expansion $(I + \epsilon)^{-1} = I - \epsilon + O(\epsilon^2)$ was carried out.
	} 
	Developing the parentheses, we see a $-R^{-1}$ term appears, canceling the $R^{-1}$ term (second term in the right hand side), and we end up with:
	$$
	\tilde J_{k,k}^{++} = \tilde J_{k,k} + \tilde J_{k,l} + \tilde J_{l,k} + \tilde J_{l,l} + O(R^2)
	$$
	And finally we can make $R$ tend to zero (i.e., precision of the virtual measurement to infinity) to recover \eqref{eq:uncloning}. Interestingly, the action of ``informing two states they are clones'', easily encoded in the covariance form, cannot be made in the information form, due to the infinite $R^{-1}$ terms. On the other hand the sequence  ``''informing two states they are clones, and then keeping only one'' works out beautifully in the information matrix form.

	\subsubsection{Final fusion}
	The final fusion is carried out for each step as in standard BIFM, according to \eqref{eq:final_upd}. The full pipeline is summarised in Algorithm \ref{alg:SC-BIFM}.

	\begin{algorithm}[t]
		\KwIn{$P_0, (u_i)_i, (Q_i)_i, (Z_k)_k, (R_k)_k$\;}
		\SetKwBlock{Forward}{Forward  pass}{end}
		\Forward{
			\nl Set $\tilde X_0 = Z_0$, $\tilde P_0 = P_0$ \;
			\SetKwBlock{ForLoop}{For $k < N$ do}{end}
			\ForLoop{
				\nl Compute $\tilde X_{k+1}$ and $\tilde P_{k+1}$, based on \eqref{eq::augmented_prop}, \eqref{eq::augmented_obs}, using the Kalman equations  \eqref{prop:::eq}, \eqref{up:::eq}\;
			}
		}
		\SetKwBlock{Backward}{Backward  pass}{end}
		\Backward{
			\nl Set $\tilde y_N = 0$, $\tilde J_N = 0$, $k=N$ \;
			\SetKwBlock{BackForLoop}{For $k > 0$ do}{end}
			\BackForLoop{
				\nl Compute $\tilde y_{k-1}$ and $\tilde J_{k-1}$, based on \eqref{eq::augmented_prop}, \eqref{eq::augmented_obs}   using the information form equations  \eqref{eq:backward_prop}, \eqref{eq:backward_upd} \;
			}
		}
		\SetKwBlock{FusionForLoop}{For $k \leq N$, do}{end}
		\FusionForLoop{
			\nl Compute the augmented solution $\tilde{X}^*_{I_k}$ based on \eqref{eq:final_upd} and extract only the state variables $X_k^*$ \;
		}
		\KwOut{$\chi=(\tilde{X}_k^*)_{k \leq N}$ which is the exact solution to the linearised optimization problem \eqref{eq::log-likelihood22}\;}
		\caption{SC-BIFM}
		\label{alg:SC-BIFM}
	\end{algorithm}

	\begin{rem}
		Alternatively, at step 5 of Algorithm \ref{alg:SC-BIFM}  one could first  extract simple state variables $X_k$ from augmented ones $\tilde X_k=X_{I_k}$, and then perform fusion \eqref{eq:final_upd} with smaller matrices to save computation time. However the extraction is a marginalization, which has to be performed on the information variables $\tilde y,\ \tilde J$, which is computationally costly.  
	\end{rem}

	Note that from the final update we may also get the posterior covariances of the related variables. However, this only yields covariances of variables which are neighbors in the measurement graph, contrary to more general methods \cite{kaess2009covariance}. 
	To this respect,  note that an alternative to SC-BIFM, which does not try to maintain sparsity, is to simply not discard the created clones. Then, at the end of the forward pass, the filter outputs $P(X_0, \ldots, X_n | Z_0, \ldots, Z_n)$, that is, the   solution we pursue.

	\begin{figure}
		\includegraphics[width=0.5\textwidth]{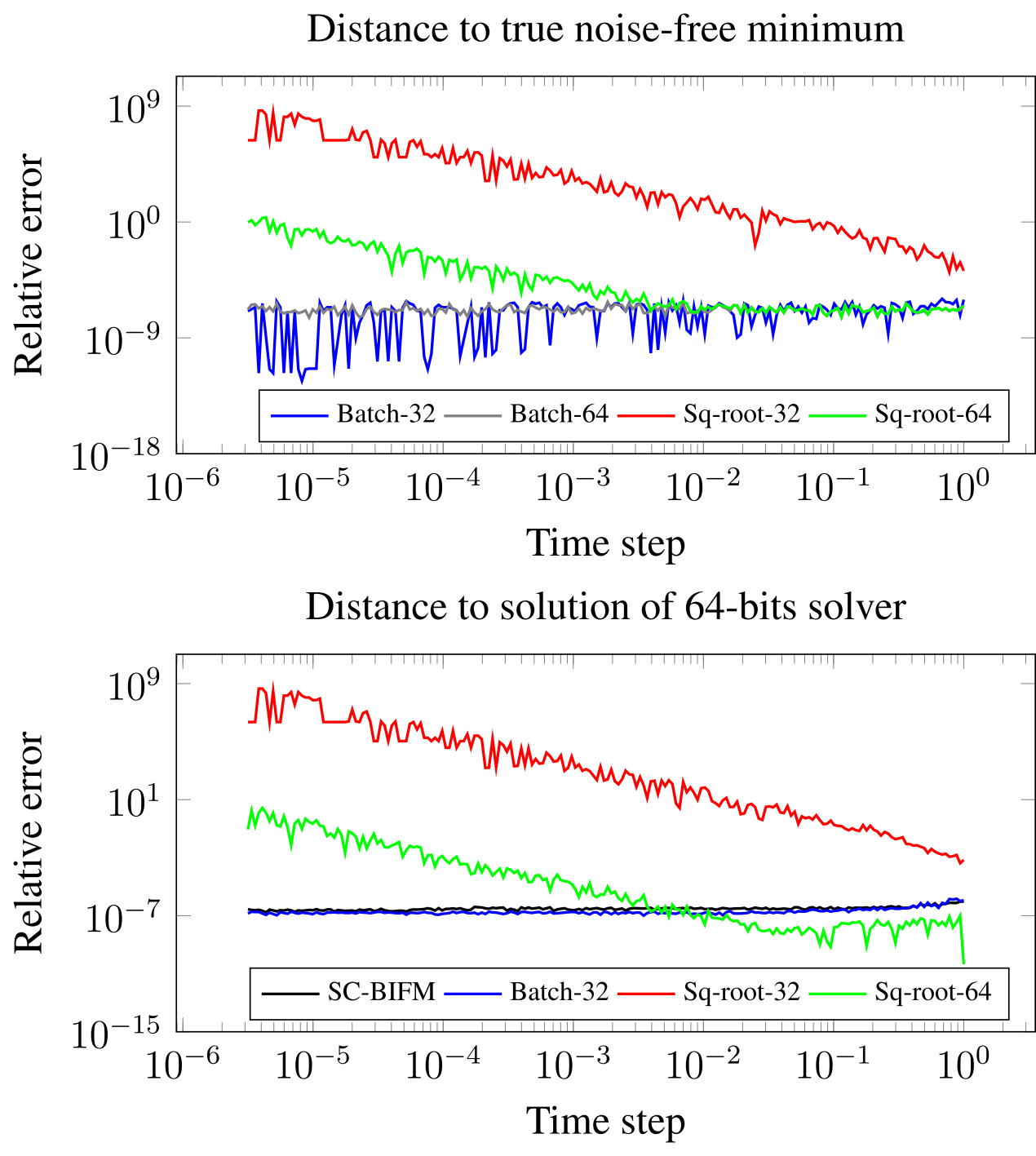}
		\caption{Top plot: distance between the solutions of various solvers and the true minimum of the simple numerical example of Section  \ref{sec:toy_example} in the process noise free case. Bottom plot:  average Monte-Carlo distance to the solution of the 64-bits robust batch solver (which systematically finds the optimum) for various solvers in the presence of process noise. In both cases performances are plotted against   the discretization time step $dt$  in \eqref{ciudad:eq2}. Standard solvers based on the information form (Sq-root-32 and Sq-root-64) degrade rapidly when $dt$ is taken small, whereas the ones we propose (Batch and SC-BIFM) do not degrade, even with single precision (32 bits).}
		\label{fig:toy_example}
	\end{figure}
	
	\section{Numerical illustration of the methodology}
	\label{sec:toy_example}
	
	We illustrate numerically the  behavior of each method using a linear  simplified navigation example. Consider a body equipped with a biased accelerometer moving along a  horizontal line. By letting $p(t)$ denote its position, $v(t)$ its velocity, and $b$ a  static  accelerometer bias, the noiseless dynamical motion equations read
	$$
	\dot p=v,~\dot v=(u+b),\dot b=0
	$$
	with $u$ the measurement reported by the accerelorometer. We suppose morevoer that at discrete time instants, using for instance vision, the system is able to measure relative displacement with respect to a past position. In discrete time, adding some process and measurement noise, and following the structure of the graph  of Figure~\ref{fig:graph}, the state $X = (b, v, p)$ is approximately governed by
	\begin{align}
	X_{k+1} &= 
	\begin{pmatrix}
	1 & & \\
	-dt & 1 & \\
	& dt & 1
	\end{pmatrix}
	X_k + \begin{pmatrix} 0 \\ 1 \\ 0 \end{pmatrix}u_k+w_k \label{ciudad:eq}\\
	Z_3 &= \begin{pmatrix} 0 & 0 & 1 \end{pmatrix} (X_3 - X_0)+V_k\label{ciudad:eq2}\\
	Z_4 &= \begin{pmatrix} 0 & 0 & 1 \end{pmatrix} (X_4 - X_2)+V_k
	\end{align}
	{  This model is a simplified version of the system dynamics used in inertial navigation. Usually, position integration with respect to the speed is considered exact on a single time step, however IMU outputs are given at a much higher rate than the observations, and are therefore preintegrated between two states \cite{forster2016preintegration}. This leads to a propagation factor with full-rank (although low eigenvalues) covariance matrix $Q$. We take the length of the trajectory to be $N=4$. Observation noise $V_k$ has variance $\sigma_z$. The process noise $w_k$ has diagonal covariance   $\mathrm{diag}(\sqrt{dt} \sigma_b,\sqrt{dt} \sigma_{acc},\sqrt{dt} \sigma_{int})$, which represent the bias random walk, the accelerometer and the integration uncertainty respectively.
	}
	Their magnitude reflects typical characteristics of high-grade inertial sensors:
	\begin{align}
	\sigma_b = 1e-3, \ \sigma_{acc} = 1e-2, \ \sigma_{int} = 1e-3, \label{g1}
	\end{align}with   initial standard deviations and observation noise \begin{align}
	\sigma_b^0 = 1e-2, \ \sigma_v^0 = 1, \ \sigma_p^0 = 1, \sigma_z=0.1. \label{g2}
	\end{align}
	
	We compare the following solvers:
	\begin{itemize}
		\item The standard square-root information solvers based on Cholesky (or   QR) factorisation of Section \ref{sqrt:sec}, with precision of 64 (Sq-Root-64) and 32 bits (Sq-Root);
		\item The robust batch solver (``Batch solver'')   of Section \ref{sec:batch_solver}, with precision of 64 (Batch-64) and 32 bits (Batch 32);
		\item The proposed SC-BIFM solver in 32 bits (only).
	\end{itemize}
	
	\subsection{Ideal case numerical experiment}
	
	{First, we study the impact of conditioning on the solvers in the noise free case, that is where the ground truth is described by \eqref{ciudad:eq}, \eqref{ciudad:eq2} with $w_k = 0$ and $V_k = 0$. The state is initialised at the ground-truth, so that we can provide the true minimum as ground truth, by simply integrating \eqref{ciudad:eq}. The estimation here is made relative to the initial position $p_0$, which is naturally considered fixed and is removed from the state. The effect of the time step $dt$ on the numerical behavior of the various solvers is displayed on Figure \ref{fig:toy_example}, top plot. Indeed, $dt$ is tightly linked with the system's conditioning, since the propagation covariance varies with its square-root.
	}

	We can see that, as expected, the   solvers based on information matrix degenerate as the time step becomes too small, while the others maintain a stable relative error, in both 32 and 64 bits. SC-BIFM is not displayed here, as it exactly follows the true state owing to perfect initialization in the forward pass.  
	
	\subsection{Numerical experiment with low noise} Monte-Carlo simulations were run with noise turned on. Since the 64 bits batch robust solver actually managed to find the true solution of the linear least-squares problem in all cases, it was taken as the reference for the comparison of the remaning solvers. 200 sets of measurements were randomly generated with noises \eqref{g1}-\eqref{g2}. The distance between the solution of each solver and the ground-truth are computed at each run, and the average distances are shown in Figure \ref{fig:toy_example}, bottom plot. In the noisy case, we also see that the standard solvers perform equivalently or slightly better than the robust ones at large time steps, but degrade rapidly for smaller ones .
	These results clearly indicate that one should be careful when designing least-squares-based estimators for accurate sensors.
	
	\begin{rem}
		The degeneracy of the square-root filters comes from the fact that they are based on QR factorisation. Using an SVD   to invert the system proves to be much more robust. However, it does not maintain sparsity and is therefore not desirable in the context of factor graph navigation and SLAM.
	\end{rem}



	\section{Real world experiment}
	\label{sec:data_joint_lab}
\begin{figure}
		\centering
		\includegraphics[width = 0.4\textwidth]{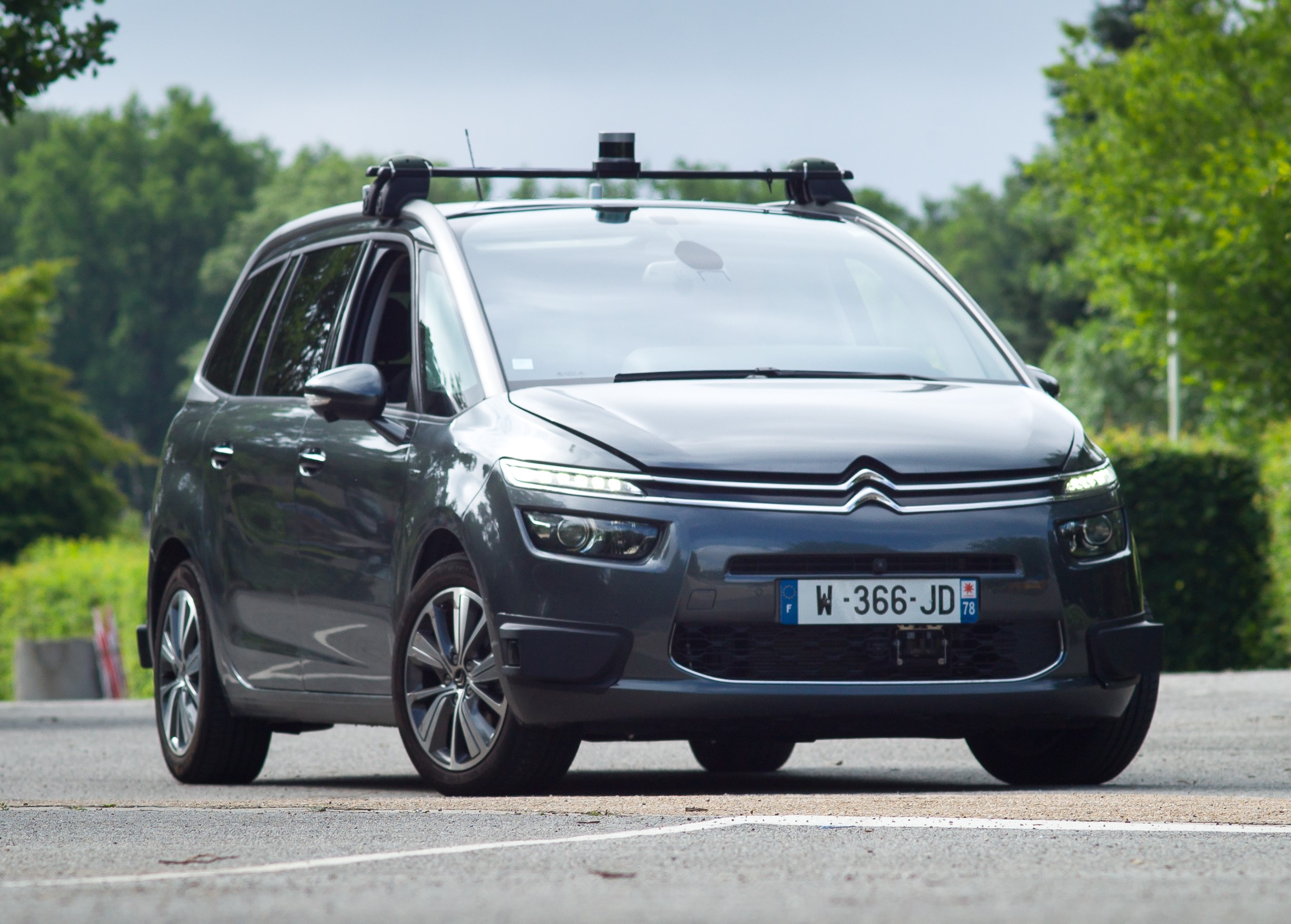}
		\caption{Car used in the experiments. }
		\label{fig:car}
	\end{figure}

	This section deals with a real high precision inertial-LiDAR odometry problem  in the context of autonomous cars.  We conducted our own experiments, using the experimental  car of Safran shown in Figure \ref{fig:car}.  In our approach,  3D LiDAR scans between keyframes were first processed to obtain relative transformations using scan matching algorithms. This imposes observations between pairs of state variables at different times of the form \eqref{eq::obs2}, which are then fused with  inertial measurements.  Combining pose from relative scans with odometry through graph optimization has been pursued in various works, see for example  \cite{mendes2016icp}. However,  combining relative scans with IMU has been less studied, and to our knowledge pose graph SLAM using LiDAR and high grade military (or ``tactical'')  precision IMUs has never been done before. This is the main originality of the following experiments, along with a suitable optimization approach to handle such levels of precisions that lead to ill-conditioned normal equations. 

	\subsection{Considered models}
	We use the standard IMU preintegration framework, see \cite{forster2016preintegration} to define our propagation factors. The LiDAR scans are aligned independently to obtain relative poses.
	{ 
		Let $X_k = (R_k, v_k, p_k, b^\omega_k, b^a_k) $ represent the state variables: $R_k, v_k, p_k, b^\omega_k, b^a_k$ are the orientation, velocity, position and biases. This leads to the following dynamics, see e.g., \cite{forster2016preintegration}:
		\begin{equation}
		\begin{aligned}
		R_{k+1} &= R_k \exp_{SO(3)}(dt\ \omega_k - b_k^\omega + w_k^\omega) \\
		v_{k+1} &= v_k + dt\ (R_k (a_k - b_k^a + w_k^a) + g) \\
		p_{k+1} &= p_k + dt\ v_k \\
		b^\omega_{k+1} &= b^\omega_k + w_k^{b_\omega} \\
		b^a_{k+1} &= b^a_k + w_k^{b_a},
		\end{aligned}\label{fullNL}\end{equation}
		where $w_k^\omega, w_k^a, w_k^{b_\omega}, w_k^{b_a}$ are white noises, associated to the isotropic covariances of standard deviations $\sigma_\omega, \sigma_a, \sigma_{b^\omega}, \sigma_{b^a}$ respectively.

		\subsection{Experimental setup}
		
		All experiments were performed using the car shown in Figure \ref{fig:car}, which is equipped  with a high-grade IMU from the company Safran Electronics and Defense and a Velodyne VLP32C. The car drove along an 11$km$ long loop in a suburban area, including medium-speed (90$km/h$) portions and round-abouts. The IMU provided data at 100 $Hz$, for which we set, according to the specification given by the manufacturer:
		\begin{align*}
		P_0 &= \text{diag}((\sigma^0_R \ I_2, 0), \sigma^0_v \ I, 0, \sigma^0_{b^\omega}\ I, \sigma^0_{b^a}\ I), \\
		\sigma^0_R &= 1\ ^\circ, ~
		\sigma^0_v = 100\ m/s,  ~
		\sigma^0_{b^\omega} = 0.01\ ^\circ/s, \\
		\sigma^0_{b^a}  &= 0.05\ m/s^2,  
		\sigma_\omega  = 0.1\ ^\circ/h \\
		\sigma_a &= 0.0015\ m/s^2 ~ (0.15 mg),  ~
		\sigma_{b^\omega}  = 2.5e-5\ ^\circ/s/\sqrt{s} ,\\  ~
		\sigma_{b^a}  &= 8e-4\ m/s^{-2}/\sqrt{s}.
		\end{align*}
		From the acquired 3D laser scans, relative transforms were obtained at 4 Hz.  Scan matching algorithms such as the well-known iterative closest point (ICP) \cite{besl1992ICP}, or more recent methods \cite{zhang2014LOAM, deschaud2018IMLS}, return relative orientation and translation. As the   gyro only drifts of $0.1$ degree in 1 hour, the level of uncertainty associated to the relative rotation between LiDAR scans computed by the ICP is much higher than  the gyro's uncertainty. On the other hand, we estimated relative translations between scans computed using the ICP have an accuracy of order   $10$ cm (and   the measurement covariance matrix $R$  is set accordingly), which is  more precise than the relative displacement obtained via a double integration of the accelerometer.   As a result we use relative translations as an observation between pairs of states \eqref{eq::obs2}. 
		
		\subsection{Implementation}
		A standard smoothing on-manifold approach was used to carry out the estimation to cope with the nonlinear structure of rotations. A 5 seconds sliding window approach was adopted, in which   states  are added each time a relative translation measurement is received. Once the number of estimated states exceeds the window's size, the oldest state is marginalised out.    The Gauss-Newton strategy described in Section \ref{sec:smoothing} was used, with one descent iteration carried out each time a measurement was acquired. Different solvers were implemented for comparison purposes. To focus on them, the rest of the pipeline was fully treated in double precision. All algorithms were implemented in Python, based on ``Numpy'' and ``Scipy'' built-in functions. In the following we detail the specifics of the implementation of the IMU preintegration and the solvers.
		
		\subsubsection*{Computing the uncertainties of the prior and the IMU preintegration}
		The preintegration of IMU increments was done following \cite{forster2016preintegration}. The associated uncertainty, however, was computed by propagating the square-root of the associated covariance \cite{wadehn2016square}. Likewise, the uncertainty of the prior was computed in square-root form. This was needed to avoid some numerical issues, and was useful for feeding the Square-Root solvers with the most     accurate uncertainty estimation.
		
		\subsubsection*{SC-BIFM}
		It was implemented according Algorithm \ref{alg:SC-BIFM}. The covariance matrices $P_0$ and $Q$ were retrieved by simply squaring the triangular matrices stored for the Prior and the IMU.
		
		\subsubsection*{Square-Root Smoothing}
		Square-Root smoothing was achieved  using the QR decomposition of $\tilde{A}$, based on its ``Numpy'' implementation. Then the resulting linear systems were solved thanks to the ``Scipy'' package.
		
		\subsubsection*{Robust Batch Approach}
		As for SC-BIFM, the Batch solver was fed with full covariance matrices (i.e., not square roots). The solution was computed using the ``Scipy.sparse'' package, and especially its ``spsolve'' method to invert the ensuing square linear systems.
	} 
	
	\subsection{Experimental comparison of linear solvers}
	\label{sec:real_comparison}

	Three linear least squares solvers are compared, in their 32- and 64-bits floating point formats:
	\begin{itemize}
		\item The standard square-root information solvers based on Cholesky (or   QR) factorisation of Section \ref{sqrt:sec}, with precision of 64 (Sq-Root-64) and 32 bits (Sq-Root-32);
		\item The robust batch solver (Batch solver) based on reformulation of Section \ref{sec:batch_solver}, with precision of 64 (Batch-64) and 32 bits (Batch 32);
		\item The proposed SC-BIFM solver in 64 and 32 bits.
	\end{itemize}
	As  noticed in the simple numerical example of Section \ref{sec:toy_example}, the best method in terms of accuracy (but not in terms of complexity) is the Batch formulation  of Section \ref{sec:batch_solver} which solves the linear least squares while avoiding ill-conditioned related issues. As a result, to provide all  solvers meaningful linearization points, we performed fusion betwen IMU and relative pose measurements based on scan matching using the Batch solver with 64 bits  based double precision.

\begin{figure}
		\includegraphics[width=0.5\textwidth]{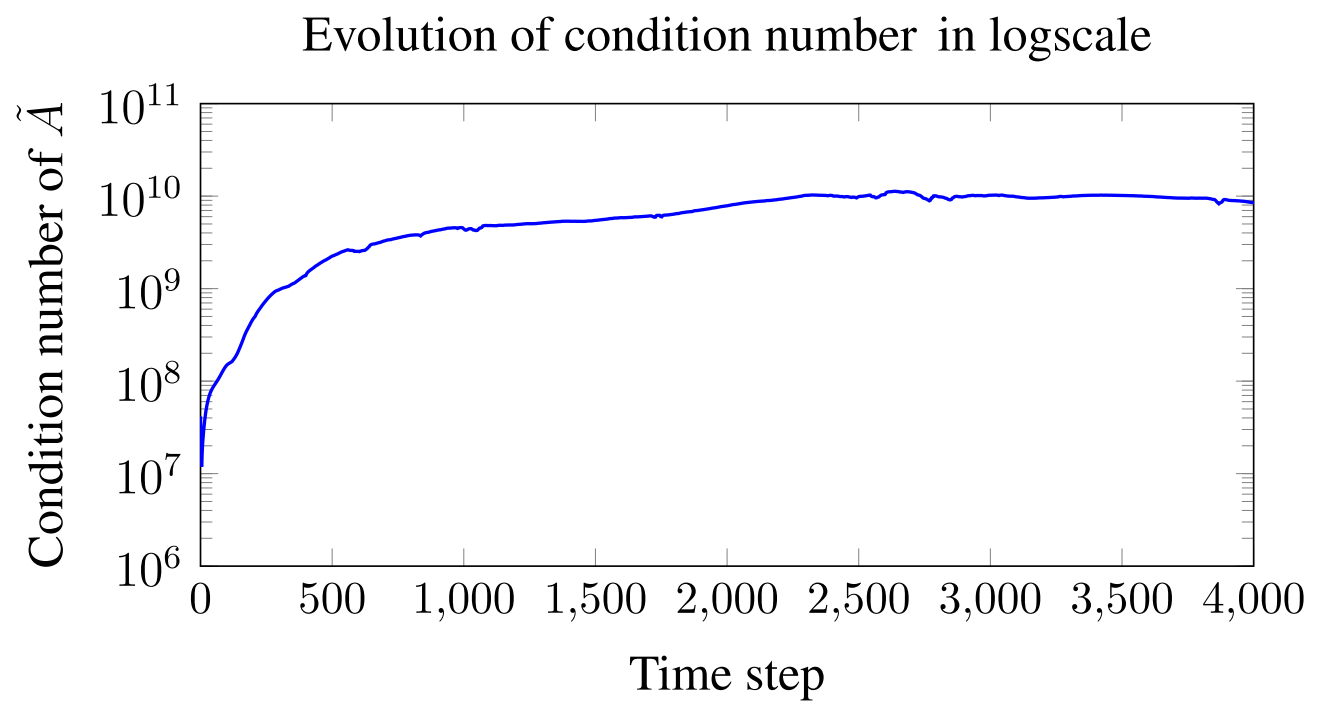}
		\includegraphics[width=0.5\textwidth]{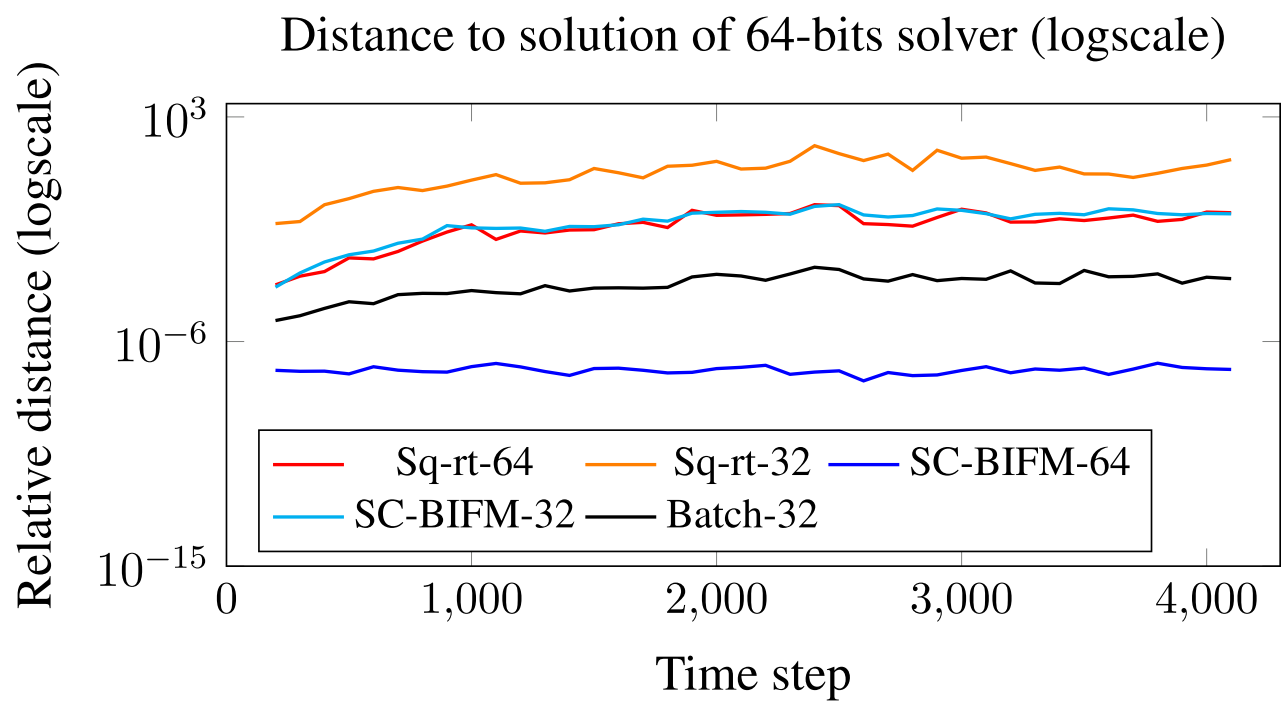}
		\caption{Top :  Evolution of the condition number of $\tilde{A}$ during the trajectory, in log scale; Bottom : Distance to Batch 64 considered herein as ground truth, in log scale. A moving average on 100 steps was performed to improve readability. There is a clear link between the increase of ill-conditioning and the accuracy degradation of the single precision solvers. However, the proposed SC-BIFM systematically beats its square-root information counterpart described in Section \ref{sqrt:sec}. Notably it is remarkable SC-BIFM achieves comparable results with 32 bits single-precision as square root with 64 bits double-precision implementation. }
		\label{fig:real_cond}
	\end{figure}

	Each time step where a GN iteration was carried out, the condition number of the weighted jacobian $\tilde{A}=\Sigma^{-1/2}A$ associated to the full normal equations \eqref{eq::inverse}, and all the solvers were applied to the corresponding linear least squares system. Figure \ref{fig:real_cond}, shows the evolution of the condition number in log-scale, and the distance between the solution of each of the single precision solvers with respect to that of the Batch-64, which is considered herein as ground truth.  
	
	\subsubsection*{Accuracy} The methods rank as follows. The Batch approach we proposed in  Section \ref{sec:batch_solver} runs first,  then the proposed SC-BIFM, and finally the square root resolution of full normal equations. The fact that Batch beats SC-BIFM may be due to our taking  advantage of the existing libraries to solve the linear systems that involve $A_1^{-1}$, whereas SC-BIFM uses no such library owing to its different nature and implementation based on Kalman smoothing, and thus had to be coded entirely from scratch.
	
	\subsubsection*{Execution time} Average computation times of the various solvers are displayed in Table \ref{tab:time_and_dist_results} are competitive with the Sq-Root, in both precision formats. Batch solvers are slightly slower, which was expected because of the $A_1^{-1}$ term to be computed. This is on the other hand encouraging for SC-BIFM, which is based on an academic code developed from scratch and thus does not  yet benefit from the same level of code optimisation as the other methods that use bricks such as  Numpy built-in $QR$ decomposition. { Theoretically, our code runs in $O(m^3 n)$, where $n$ is the number of variables and $m$ the dimension of the largest $X_{I_k}$ related to the number of clones. It is difficult to go further into complexity analysis owing to the clones, but we  believe its complexity is closely related to the complexity of sparse Cholesky or QR decompositions, which varies with the filling of the $R$ factor. }
	
	\begin{table}[]
		\centering
		\begin{tabular}{|c|c|}
			\hline
			Solver & \thead{Average computation time ($s$)}  \\
			\hline
			Sq-Root 64 & $0.021$   \\
			\hline
			Sq-Root 32 & $0.010$  \\
			\hline
			Batch 64 & $0.029$  \\
			\hline
			Batch 32 & $0.024$    \\
			\hline
			SC-BIFM 64 & $0.020$  \\
			\hline
			SC-BIFM 32 & $0.016$   \\
			\hline
		\end{tabular}
		\caption{Average computation times for the inversion of the linear systems over trajectory. Computations were made on a standard laptop with Intel i5-5300 2.3 GHz CPU}
		\label{tab:time_and_dist_results}
	\end{table}

	\subsection{Experimental comparison of corresponding localization algorithms}
	\label{sec:real_est_single}
	Following  the full pipeline of iterative linearization procedure of factor graph based SLAM (or navigation) recapped in Section \ref{sec:smoothing}, we computed state variable estimates for the  nonlinear system \eqref{fullNL} using the various solvers.

\begin{figure*}
		\centering
		\includegraphics[]{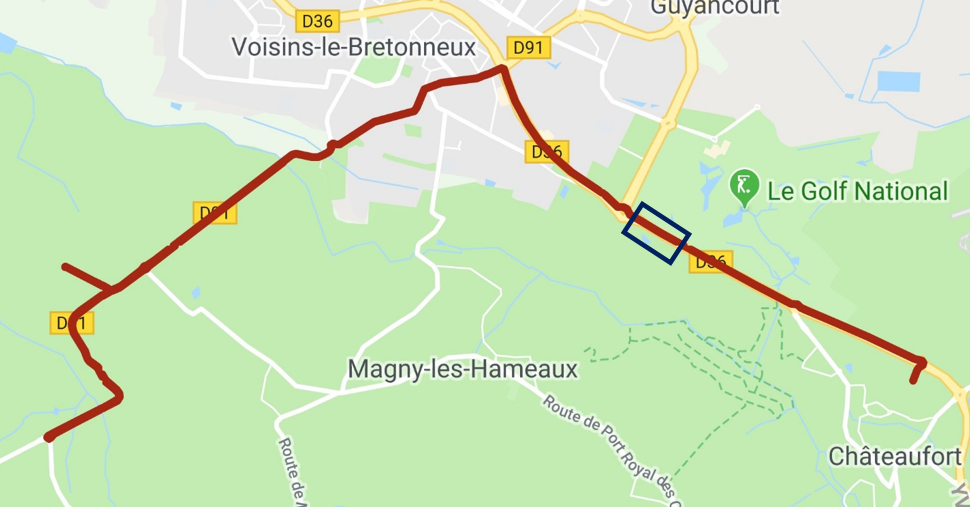}
		\begin{minipage}{.49\linewidth}
			\includegraphics[width = .95\textwidth]{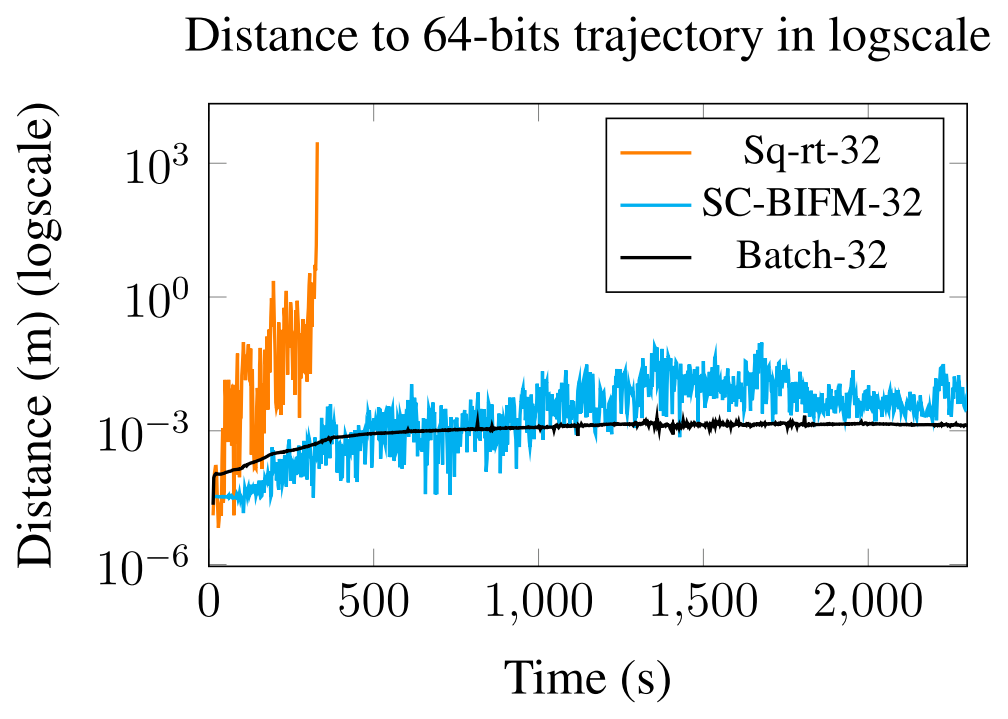}
		\end{minipage}
		\begin{minipage}{.49\linewidth}
			\includegraphics[width = .95\textwidth]{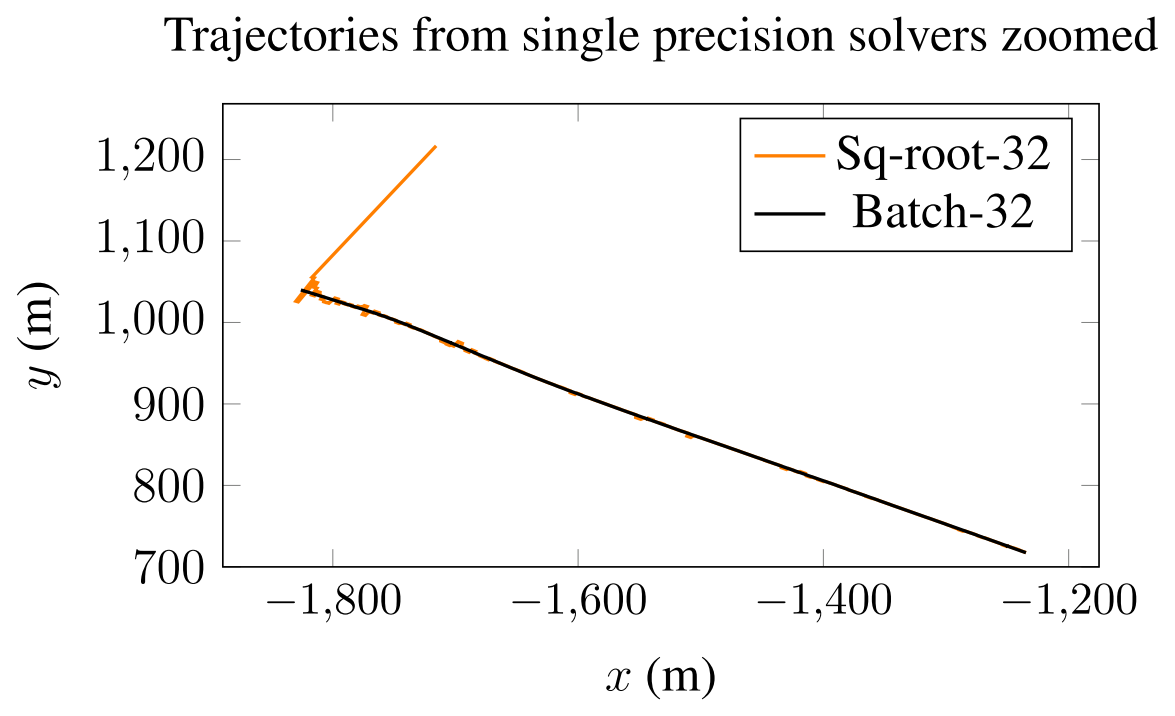}
		\end{minipage}
		\caption{Top : 11 km long trajectory followed by the car, taken from Google Maps.   The bottom right image displays a zoom on the  zone inside the black box,  and  shows the trajectories estimated by Sq-root-32 and Batch-32. We see Sq-root-32 diverges whereas Batch-32 properly follows the true trajectory. Both proposed single precision solvers, SC-BIFM-32 and Batch-32, managed to stay stable during the whole experiment, as shown in the bottom left picture. More generally, the deviations observed are consistent with the results reported in Table \ref{tab:time_and_dist_results2}.}
		\label{fig:trajectories}
	\end{figure*}

	Table \ref{tab:time_and_dist_results2} displays the maximum distance to Batch 64 estimate (considered as optimal) in terms of position discrepancy of the car. The results confirm what could be anticipated from the linear solvers comparison above. Moreover,   using single-precision Sq-root-32, the estimation could not even be carried out until the end, as it diverged after about an eighth of the trajectory, see Figure \ref{fig:trajectories}. SC-BIFM-64 is as expected the closest to Batch-64, followed by Batch-32, while Sq-root-64 and SC-BIFM-32 show similar behaviors.  The latter feature is remarkable, though, as it shows he proposed algorithm SC-BIFM may achieve good results in single precision on a real application. 
	
	\subsection{Conclusions regarding real   experiments}
	
	The first merit of the conducted experiments is to prove that the problem of ill conditionned information matrix for pose graph SLAM or Kalman smoothing for navigation may arise when using highly precise inertial sensors (such as cost effective high precision IMU from Safran Electronics and Defense). Given the progress made in the field of inertial sensors over the past decades, we can anticipate performance will keep increasing, and cost decreasing. We showed the resulting problems can even lead to divergence of the localization estimate based on the standard Cholesky resolution of linear least squares (Sq-Root 32) when implemented with 32 bits single-precision. This is an important point, as  most industrial grade inertial navigation embedded systems use 32-bit precision. In this respect, we see the two solutions we proposed in this paper are satisfactory and achieve good performance in single precision. 
	Moreover, the proposed novel SC-BIFM method based on Kalman smoothing with stochastic cloning seems promising, as its complexity is reasonably low since {  only matrices of a limited size need to be inverted, as opposed to the proposed Batch solver.
	}

	\begin{table}[]
		\centering
		\begin{tabular}{|c|c|}
			\hline
			Solver & \thead{Maximum distance to Batch-64 ($m$)} \\
			\hline
			Sq-Root 64 & $0.09$ \\
			\hline
			Sq-Root 32 &  $\infty$\\
			\hline
			Batch 64 &  $0$\\
			\hline
			Batch 32 &  $0.01$ \\
			\hline
			SC-BIFM 64 & $7e{-4}$\\
			\hline
			SC-BIFM 32 & $0.1$ \\
			\hline
		\end{tabular}
		\caption{Maximum distance to the Batch-64 trajectory for the various solvers.}
		\label{tab:time_and_dist_results2}
	\end{table}
	
	\section{Conclusion}
	In this paper we highlight a shortcoming of the standard solvers used to invert the linear least-squares problems appearing in smoothing methods for sensor fusion, and especially inertial navigation. It was shown that these methods are not adapted to overly low propagation noise, and can degenerate, for instance in single precision. We build on the fact that some formulations of Kalman smoothers avoid this numerical issue and propose a novel sparsity-maintaining solver for a large class of systems, using stochastic cloning. It exhibits much higher robustness than standard solvers in the face of the mentioned numerical difficulties.

	Research to achieve numerical efficiency comparable to the one of existing solvers is still needed, as they benefit from decades of numerical algorithms, along with major recent breakthroughs from the SLAM community.   Another route for future research is to come up with an incremental formulation of SC-BIFM.   Indeed, the forward part is easily continued when a new measurement is available, but the backward information filter seems to have to be recomputed from scratch each time. We anticipate that there might be links with the Bayes tree used in iSAM2 \cite{kaess2012iSAM2}, for instance, which could help resolving this issue.

	\section{Acknowledgements}
	This work is supported by the company Safran through the CIFRE convention 2016/1444.


	\bibliography{biblio}
	\bibliographystyle{plain}

\end{document}